\documentclass{aa}

\usepackage{graphicx}
\usepackage{txfonts}
\usepackage{graphicx}
\usepackage{amsmath}
\usepackage[version=4]{mhchem}
\usepackage{siunitx}
\usepackage{pdflscape}
\usepackage{longtable,tabularx}
\usepackage{float}
\usepackage{multirow}
\usepackage{multicol}
\usepackage{url}
\usepackage{comment}
\usepackage{xcolor}
\usepackage{adjustbox}
\usepackage{caption}  
\usepackage{subcaption}

\begin{document}

\title{Near Earth stream decoherence revisited: the limits of orbital similarity}

\titlerunning{Stream Decoherence}

\author{P.M. Shober \inst{1,2}  \and
        A. Courtot \inst{3,1} \and
        J. Vaubaillon \inst{1}
        }

\institute{
IMCCE, CNRS, Observatoire de Paris, PSL Universit\'e, Sorbonne Universit\'e, Universit\'e de Lille 1, UMR 8028 du CNRS, 77 av. Denfert-Rochereau 75014 Paris, France  \quad \email{patrick.shober@obspm.fr}
\and 
Space Science and Technology Centre, Curtin University, GPO Box U1987, Perth WA 6845, Australia
\and
 ESA Space Environments and Effects Section (TEC-EPS), ESTEC, the Netherlands
}

\date{Received xx 2024; accepted XXX 2024}

\abstract
   {Orbital similarity measures, such as the D-values, have been extensively used in meteor science to identify meteoroid streams and associate meteorite falls with near-Earth objects (NEOs). However, the chaotic nature of near-Earth space challenges the long-term reliability of these measures for stream identification, and the increasing size of our fireball, meteorite fall, and NEO databases make random associations more common. Despite this, many researchers erroneously continue to use orbital similarity beyond its inherent limits.}
   {We aim to assess the statistical significance of using orbital similarity measures for identifying streams of meteoroids or asteroids and explore the implications of chaotic dynamics on the long-term coherence of these streams.}
   {We employed a Kernel Density Estimation (KDE) based method to evaluate the statistical significance of orbital similarities within different datasets. Additionally, we conducted a Lyapunov characteristic lifetime analysis and simulated 300 fictitious meteoroid streams to estimate the decoherence lifetimes in near-Earth space. The orbital similarity was determined using the D$_{SH}$, D', and D$_{H}$ orbital similarity discriminants. Clustering analysis relied on a Density-based spatial clustering of applications with noise (DBSCAN) algorithm.}
   {Our analysis found no statistically significant streams within the meteorite fall, fireball, or USG impact datasets, with orbital similarities consistent with random associations. Conversely, 12 statistically significant clusters were identified within the NEO population, likely resulting from tidal disruption events. The Lyapunov lifetime analysis revealed short characteristic lifetimes (60–200 years) for orbits in near-Earth space, emphasizing the rapid divergence of initially similar orbits. Meteoroid stream decoherence lifetimes ranged from $10^4$ to $10^5$ years, aligning with previous studies and underscoring the transient nature of such streams.}
   {The rapid decoherence of meteoroid streams and the chaotic dynamics of near-Earth orbits suggest that no reported stream or NEO associations of meteorites or fireballs are statistically significant according to orbital discriminates. Many are likely coincidental rather than indicative of a true physical link. However, several statistically significant clusters found within the NEO population are consistent with a tidal disruption formation. This contrast and lack of statistically significant associations amongst the impact datasets is likely due to the fireball databases being 2 orders of magnitude smaller than the NEO database and the higher intrinsic uncertainties of fireball observation derived orbits.}


\maketitle

% \todo{General comment by JV: this article was first entitled: "Kill $D_{SH}$" and the point I wanted to make is that this criterion should NOT be used (in general, not only in meteor sciences). So why are we using it in this paper? I think we should 1) add a section to fully explain/recall why $D_{SH}$ should never have been used in the first place (from e.g. Ariane's work) and 2) remove all references to $D_{SH}$ in the method and results. Otherwise it doesn't make sense at all. Well ok maybe what happened is that Patrick had a broader view from what I first envisioned and (my bad...) I didn't check the amnuscript before now... Sorry Patrick if this bothers you... Let's talk about this when you're back in Paris}

\section{Introduction}\label{sec:intro}
Stream identification and differentiation from the sporadic background are crucial components of meteor science. The International Astronomical Union (IAU) actively updates a detailed registry\footnote{\url{https://www.ta3.sk/IAUC22DB/MDC2022/}} of meteor showers and their originating celestial bodies. This registry is frequently augmented by findings from new meteor or fireball observation surveys, highlighting the evolving landscape of this scientific domain \citep{jenniskens2009report,kornovs2014confirmation,rudawska2015independent,jenniskens2016established}. Identifying these streams relies on employing a dissimilarity measure known as the D-function. Initially introduced by \citet{southworth1963statistics}, the $D_{SH}$ is derived by comparing two groups of orbital elements. This measure, which increases as the similarity decreases, is often termed a `dissimilarity parameter.' Several adaptations have been introduced, each slightly changing the degree of influence of different elements to the D value \citep{drummond1981test,steel1991structure,jopek1993remarks,jopek2008meteoroid,jenniskens2009report}. 

Despite its extensive application, the efficacy of the D-values or orbital similarity in general has been contested, particularly its dependency on sample size and the risk of falsely identifying random groupings as actual meteor showers \citep{pauls2005decoherence,koten2014search,egal2017challenge,vida2018modelling,shober2024generalizable}. With the surge in accessible meteor orbit data from sources like video meteor networks, the demand for more reliable techniques to verify links between meteor showers and their parent bodies has intensified. This requirement is amplified by recent acknowledgments of the higher-than-anticipated uncertainty surrounding meteoroid orbital elements \citep{egal2017challenge,vida2018modelling,shober2023comparison}. To address these issues, new statistically robust methods have been proposed, such as the wavelet transform \citep{galligan2002wavelet}, application of the DBSCAN (Density-Based Spatial Clustering of Applications with Noise) algorithm to identify clusters in meteor trajectory data \citep{moorhead2016performance,sugar2017meteor} or the estimation of meteor shower false positive detections leveraging Kernel Density Estimation (KDE) \citep{shober2024generalizable}.

In spite of the ongoing challenges, at least 110 meteoroid streams\footnote{\url{https://www.ta3.sk/IAUC22DB/MDC2007/Roje/roje_lista.php?corobic_roje=1&sort_roje=0}} have been confirmed and solidly associated with primarily dust ejection from cometary activity \citep{jopek2017iau,jenniskens2020removing}. Streams of larger meteoroids and asteroids have been the subject of many studies \citep{fu2005identifying,schunova2012searching,jopek2020orbital}; however, no significant evidence supports the associations between the meteoroids and NEOs. Recently, likely clusters within the NEO population have been identified by \citet{jopek2020orbital}; however, no considerable statistical evidence has supported associations between fireball data and such NEOs. 
% Several recent studies have found fireball, USG Sensor impact data, or meteorite falls with similar orbits with NEOs; however, each contains one or multiple statistical oversights concerning statistical significance. 
If a macroscopic meteoroid stream existed, it could pose a consistent impact threat, and thus, the possibility has been continuously explored. \citet{pauls2005decoherence} was the first comprehensive study aimed to investigate whether large meteoroid streams, proposed to be responsible for the similarity of a handful of meteorite falls, could remain coherent over long enough periods to be found in the fireball and meteorite fall datasets. Initially, they explored the statistical likelihood of orbital similarity within known meteorite-dropping fireballs, comparing these occurrences against randomly generated samples of orbits.  \citet{pauls2005decoherence} performed detailed numerical simulations on three well-documented cases of meteorite falls (Innisfree, Peekskill, and Příbram) that had been suggested to belong to meteoroid streams due to their orbital similarities with other falls or fireballs. 

The objective of \citet{pauls2005decoherence} was to address the three arguments used for the existence of macroscopic meteorite-dropping meteoroid streams: (1) orbital similarity (Příbram/Neuschwanstein and Innisfree/Ridgedale) \citep{halliday1987detection,spurny2003photographic,halliday1990evidence}, (2) the clustering of fireball data \citep{mccrosky1979prairie,halliday1990evidence}, and (3) meteorite falls which had the same day-of-fall and similar chemical trace elements \citep{lipschutz1997meteoroid}. To address these arguments, \cite{pauls2005decoherence} determined how long these assumed streams could maintain their coherence before becoming indistinguishable from the sporadic background. They showed that if these three meteorites were associated with meteoroid streams formed in near-Earth space, they would become decoherent on relatively short timescales of $10^{4}$ to $10^{5}$ years. This rapid dispersion would require an extremely recent breakup of the parent body, much more recent than the cosmic ray exposure ages of the meteorites involved and much more frequently than near-Earth asteroid impact rates \citep{bottke1994collisional}. Furthermore, the statistical evidence for clustering similar orbits was demonstrated to occur just as frequently in a random sample of orbits \citep{pauls2005decoherence}.

Since these results nearly two decades ago, many further studies have tried to use the orbital similarity of meteorite falls to find parent bodies or streams. This warrants a renewed analysis of the statistical significance of clustering in the meteorite fall and fireball datasets. Since 2005, there has been an exponential growth in the number of meteor and fireball observation networks, significantly augmenting our knowledge and ability to detect smaller streams. This is exemplified by the fact that in 2005, there were 10 known meteorite falls with precise orbits, whereas in 2024, there are at least 55 in total, with 18 alone recovered in the last five years. This is primarily due to several new networks operating around the globe dedicated to recovering meteorites \citep{bland2012australian,devillepoix2020global,colas2020fripon,borovivcka2022_one}. Given this increased capacity and the recent use of orbital similarity to make grand conclusions about source regions or parent bodies, we must re-address this question of macroscopic meteoroid streams. 

% As you know, cosmic ray exposure (CRE) ages for meteorites indicate they have been detached from their parent bodies for often millions or tens of millions of years. This timescale significantly exceeds the typical coherence lifetimes ($\sim$10kyrs) of meteoroid streams in near-Earth space (Pauls & Gladman 2005) or the Lyapunov timescales (often <250 yrs). The attempt to trace back meteorite orbits over 100,000 years using orbital similarity discriminants, although intriguing, faces these fundamental challenges. This timescale is still considerably shorter than the CRE ages, and the chaotic nature of orbital dynamics over such extended periods excludes the possibility of using orbital similarity to identify parent bodies for meteorites with older CRE ages. This is a critical issue, and given our intention to continue our collaboration, we believed it was something we needed to address.

% \subsection{The $D_{SH}$ value}
% Now have a look at the fucking $D_{SH}$ formula: anything wrong with it?
% If not, go back to previous section.

% See also Courtot et al. 2023 for education purpose.

The article is organized as follows: section \ref{sec:method} describes the three methods used in this analysis.
Section \ref{sec:results} shows the results, and a discussion follows in section \ref{sec:discuss}.

% ==============================
\section{Methods}\label{sec:method}
To address the utility of orbital similarity for meteorite-dropping fireballs, near-Earth objects, and the existence of large meteoroid streams, we have used three primary methods: (1) a false-positive estimation and orbital clustering significance analysis using $D_{SH}$ \citep{southworth1963statistics}, $D'$ \citep{drummond1980meteor}, and $D_{H}$ \citep{jopek1993remarks}, (2) Lyapunov characteristic lifetime mapping of $a$-$e$-$\iota$ space, and (3) decoherence lifetime estimation of 300 fictitious streams.

\subsection{Data Collection and Reduction}
We examine several datasets in this study, including (1) 50 meteorite falls with orbits, (2) 616 potential 1\,g meteorite-dropping fireballs collected by the FRIPON, EFN, and GFO networks (350 of which are probably 50\,g droppers), (3) 3,290 FRIPON fireballs, (4) 824 EFN fireballs taken from \citep{borovivcka2022_one}, (5) 35,012 telescopically observed NEOs from the JPL HORIZONS\footnote{\url{https://ssd.jpl.nasa.gov/horizons/}} database, and (6) 310 impacts detected by USG sensors. The reliability of the USG sensor orbits has been demonstrated to be inconsistent and certainly not ideal for finding streams \citep{devillepoix2019observation}; however, the data was included as recent studies have relied on the dataset using D-discriminants. 

\subsubsection{Fireball Observations}
The fireball data used in this work are derived from several key fireball observation networks, each contributing unique observations crucial for this analysis. 

The first of these is the Fireball Recovery and InterPlanetary Observation Network\footnote{\url{https://www.fripon.org/}} (FRIPON), which is an expansive network primarily covering Europe, equipped with over 220 all-sky cameras covering 15 countries across the globe. It focuses on both detecting fireballs and recovering meteorites with its CCD video observatories. FRIPON's system is particularly designed to capture fireballs brighter than magnitude zero, providing detailed data for trajectory and orbit determination for meteorite-dropping fireballs along with some smaller, largely shower components \citep{colas2020fripon}.

We also utilized data collected by the Global Fireball Observatory\footnote{\url{https://gfo.rocks/}} (GFO). The GFO is an expansive collaboration that leverages the extensive infrastructure developed by the Desert Fireball Network (DFN) \citep{devillepoix2020global}. The partnership comprises ten partner networks, supported by 18 collaborating institutions spread over nine countries globally. There are over 100 observatories spread across the networks, with 50 covering over a third of the Australian landmass. The weather-hardened digital fireball observatories feature high-resolution DSLR cameras, all-sky fisheye lenses, and GNSS synchronized liquid crystal shutters for precise timing. Facilitating the recovery of 15 meteorites since the DFN's digital upgrade in 2013-2015, the collaboration has recovered approximately 30\% of all meteorites recovered with known orbits to date \citep{King_Winchcombe2022SciA, shober2022arpu, Devillepoix_Madura_Cave, Anderson2022ApJ}. For trajectory analysis, the GFO employs a straight-line least squares (SLLS) method combined with an extended Kalman smoother to ensure accurate velocity estimates \citep{Borovicka_1990BAICz, sansom2015novel}. This sophisticated approach integrates observational uncertainties to manage and propagate errors effectively. Such methodologies allow for the precise determination of pre-entry orbits by numerically integrating meteoroids' states beyond Earth's gravitational influence, taking into account all significant perturbative effects \citep{jansen2019comparing, shober2019identification, shober2020did}.

Lastly, the European Fireball Network (EFN) is the longest-running photographic fireball network, with the first observations taking place in 1963. It has a total coverage area of about one million square kilometers. A history spanning several decades, the EFN team has successfully facilitated the recovery of at least 13 meteorites \citep{spurny2003photographic,spurny2013trajectory,spurny2017ejby,spurny2020vzvdar}. The network primarily operates Digital Autonomous Fireball Observers (DAFOs), which are weather-proof and fully autonomous systems designed to continuously monitor the entire sky during clear weather conditions. Each DAFO utilizes DSLR cameras fitted with fisheye lenses, similar to those used by the DFN, capturing high-resolution images of meteor events. These cameras are configured to detect meteors with an absolute magnitude brighter than -2 and can perform detailed radiometric measurements for fireballs exceeding -4 in magnitude \citep{borovivcka2022_one}. EFN's detection system is adept at identifying meteoroids of mass greater than 5 grams and capable of capturing high-velocity meteoroids as small as 0.1 grams. Spread across central Europe, the network's 26 stations cover an area of approximately one million km$^{2}$, providing extensive observational data.

The dataset from EFN utilized in this study has been detailed in \citet{borovivcka2022_one} and \citet{borovivcka2022_two}. The atmospheric trajectory of each fireball is deduced using the Straight Line Least Squares (SLLS) method, which assumes a linear trajectory through space \citep{Borovicka_1990BAICz}. Unlike the DFN, EFN employs a different approach for velocity calculation, using time data projected onto the computed SLLS trajectory and fitting it with a physical model that accounts for atmospheric drag and ablation \citep{pecina1983BAICz..34..102P}. This model includes parameters for preatmospheric velocity, ablation coefficient, and mass-related properties, refined using atmospheric density models like CIRA72 or NRLMSISE-00. Manual adjustments or alternative models are used when significant deceleration occurs. Outliers were evaluated or removed to ensure data accuracy, and any systematic discrepancies between cameras were resolved before finalizing the heliocentric orbits, which are calculated with a modified version of the method described in \citet{Ceplecha_1987}, accounting for Earth's rotation and gravitational influences.

The 350 `meteorite-dropping' fireballs observed by the FRIPON, EFN, and GFO projects were identified primarily using the $\alpha-\beta$ methodology of \citet{sansom2019determining}. The $\alpha-\beta$ coefficients correspond to the ballistic coefficient and mass-loss parameter, respectively. These parameters, introduced and developed in \citet{gritsevich2007og} and \citet{lyytinen2016implications}, can be estimated for any impact event with some degree of deceleration. These parameters provide a simple yet effective way to characterize the probability of a likely fall without needing a multivariate solution or supercomputing resources \citep{sansom2019determining}. The 616 potential dropping events were identified within the FRIPON and GFO datasets using a variation of equations 7 and 8 in \citet{sansom2019determining}, using an end mass of 1\,g (616 events) and 50\,g (350 events). The EFN meteorite dropping subset was taken from \citet{borovivcka2022_one}, using only events with an end mass above 1\,g or 50\,g according to their ablation/fragmentation model. 
% Given the low minimum mass limit of 50\,g, a substantial portion of these 350 events may not have dropped a macroscopic sample; however, we believe this to be acceptable as the source regions and orbital distributions should be the same between these populations. 

% talk about JPL data 
\subsubsection{NEO Data}
The NEO data utilized in this study was taken from the Jet Propulsion Laboratory (JPL) Horizons database, which provides comprehensive ephemeris computations essential for the precise location of celestial objects over time. Accessible via NASA's Solar System Dynamics Group, the Horizons system offers a robust, highly reliable tool for generating accurate solar, lunar, and planetary ephemerides. Utilized extensively in the astronomical community, this service facilitates detailed orbital analysis by delivering key parameters such as positions, velocities, and magnitudes across various time points. These data are crucial for high-precision object-tracking, making Horizons an invaluable resource for observational planning and ongoing research in celestial mechanics.

\subsubsection{USG Sensors Detections}
The Center for Near-Earth Object Studies (CNEOS), operated by NASA’s JPL, also maintains a fireball database\footnote{\url{https://cneos.jpl.nasa.gov/fireballs/}} that catalogs detailed observations of fireballs produced by meteoric phenomena in Earth's atmosphere. This database is compiled from data collected by U.S. Government sensors, offering vital information on the trajectory, speed, penetration depth in the atmosphere, and estimated energy release of these events. The CNEOS fireball database is a critical tool for understanding the population of large meteoroids in terms of impact energy distribution, providing global coverage for detecting large impactors. This database is very useful for increasing the statistical significance of our understanding of the subset of the largest annual impactors. In total, 310 CNEOS impact events to date have velocity and peak brightness latitude, longitude, and altitude information. Unfortunately, this valuable data is intentionally degraded to maintain the secrecy around the sensitive nature of the US military's monitoring capabilities. Thus, the data released is very unreliable for orbit determinations, with velocities varying up to over 20\% and the radiant, in some cases, being 90\,$^{\circ}$ off when compared to more precise ground-based fireball observations \citep{devillepoix2019observation}. Therefore, using this data with orbital similarity functions is quite unideal and ill-advised. Despite this, we included this dataset given its extensive use in several studies. With that being said, we deliberately chose to not try to estimate the large uncertainties and to test the statistical significance of the CNEOS fireballs even if the errors were minuscule. 

\subsection{Statistical Significance of Orbital Similarity}\label{subsec:statistical_significance}

To assess the statistical significance of orbital similarities, we employed various D-functions to measure the dissimilarity between the orbital elements of celestial bodies. The functions applied included the D$_{SH}$, D$'$, and D$_{H}$, which consider different combinations of orbital parameters such as perihelion distance ($q$), eccentricity ($e$), inclination ($\iota$), argument of perihelion ($\omega$), and longitude of ascending node ($\Omega$).

The analysis involved two aspects: (1) using the methodology of \citet{shober2024generalizable} to estimate the false positive clustering rate by fitting a KDE to the concerned populations, and (2) calculating the D-values for all pair combinations within the concerned dataset and examining the distributions.  

Using the KernelDensity class from scikit-learn, as done in \citet{shober2024generalizable}, we estimated the false positive association rate and the statistical significance of alleged meteorite-dropping streams. KDE is a non-parametric technique that estimates the probability density function of a random variable from its data points. This is achieved by placing a kernel, often a Gaussian function, on each data point and aggregating these kernels to form a smooth representation of the underlying probability density function. This technique is adept at modeling unknown distributions and handling the multimodality frequently observed in sparse datasets \citep{silverman2018density}. Moreover, KDE is recognized for its adaptability in accurately estimating densities across a range of shapes, assuming the smoothing parameter is correctly chosen \citep{seaman1996evaluation}.

We used KDE to estimate the orbital distribution of the NEO population, the USG sensor impact population, and the 50 meteorite falls with orbits. This involved generating synthetic datasets from these KDEs, equal in size to the original datasets, and evaluating the number of pairs below specific D-value thresholds to estimate false positives. This approach helps us understand if observed pairings in the data could be explained by random chance rather than physical association. A Monte Carlo simulation repeated this process 500 times for the meteorite-NEO and USG sensors-NEO pairs; from this, the standard deviations of the sample means for the numbers of random associations could be calculated to understand the significance of the number of actual associations between the populations. Please refer to \citet{shober2024generalizable} for a more detailed explanation of the KDE false-positive methodology.

Additionally, we calculated all possible D-values for every pair combination within a single dataset. This extensive analysis allowed us to evaluate whether the D-value distribution followed expected patterns from random chance or exhibited clusters indicative of genuine associations. If a significant, actively occurring process was generating streams in near-Earth space, there should be an excess of small D-values within the distributions. The total number of unique combinations within an orbital dataset follows \(C=\frac{n(n-1)}{2}\), where n is the size of the dataset. In total, within this study, we have calculated the orbital similarity D-values for over $10^{11}$ unique orbital pairs between the meteorite-fireball-NEO comparisons and the Monte Carlo procedure using KDE to estimate the statistical significance of the similarities. 

\subsection{Lyapunov Characteristic Lifetimes}

The Lyapunov characteristic exponent (LCE) quantitatively measures chaos in a dynamical system. It characterizes the rate at which trajectories initially close in phase space diverge over time, indicating the system's sensitive dependence on initial conditions. This chaos is characterized by an exponential divergence. To calculate the LCE, we employ the open-source N-body simulation software, Rebound, specifically tailored for astronomical systems \citep{rebound2012A&A...537A.128R}. The solar system model included the Sun and the eight major planets with initial conditions sourced from JPL Horizons\footnote{\url{https://ssd.jpl.nasa.gov/horizons/}}. The 'whfast' integrator was utilized for its efficiency with near-Keplerian orbits, with a time step set to 0.01 years. 

Orbits were non-uniformly generated across a parameter space defined by semi-major axes derived from the Tisserand parameter relative to Jupiter, with semi-major axis values ranging from 0.5 to 3.5\,au, eccentricities ranging from 0 to 0.95, and an inclination of 0, 10, or 20 degrees. The non-uniform distribution of particles was chosen to concentrate the particle density in chaotic regions and optimize the resolving process. In total, 289,640 particles at 0$^{\circ}$ inclination, 287,352 at 10$^{\circ}$ inclination, and 353,470 particles at 20$^{\circ}$ inclination were integrated for the map constructions (Fig.~\ref{fig:lyapunov_maps}). A lower inclination range was chosen as the asteroidal population is nearly all within this range of values. The integration time for each simulation was set to 20,000 years, consistent with other studies \citep{tancredi1995dynamical,tancredi1998chaotic,shober2024comparing}, and the Lyapunov time was calculated as the inverse of the Lyapunov exponent. This integration time was optimal for this system based on previous studies \citep{tancredi1995dynamical,tancredi1998chaotic,shober2024comparing}.

The Rebound library integrates the dynamical equations of motion alongside the variational equations, following the methodology described by \citet{cincotta2000simple}. These variational equations describe how infinitesimal perturbations to the system's state vector (position and velocity) evolve over time. A description of this can be found within the documentation\footnote{\url{https://rebound.readthedocs.io/en/latest/chaos/}} of Rebound. The integration process involves the following steps:

\begin{enumerate}
    \item Numerically solving the N-body system's equations of motion using an integrator like WHFast, IAS15, or SEI.
    \item Simultaneously integrating the variational equations to track how perturbations in initial conditions evolve.
    \item Applying orthonormalization regularly (e.g., Gram-Schmidt process) to the deviation vectors to maintain numerical stability and prevent scaling issues.
    \item Estimating the Lyapunov exponent from the average exponential rate of divergence of these orthonormalized vectors, normalized by the integration time.
\end{enumerate}

This method enables efficient and accurate estimation of the Lyapunov characteristic lifetime, which is vital for understanding the stability and evolution of meteoroid streams under gravitational influences. For a detailed description of the integration methods and theoretical background, refer to the work by \citet{cincotta2000simple} and \cite{rebound2012A&A...537A.128R}.

\subsection{Meteoroid Stream Decoherence Lifetimes}

Meteoroid stream decoherence lifetimes were determined by simulating the dynamical evolution of stream-like groups of particles under the solar system's gravitational influences. The simulations were also performed using the N-body integration software REBOUND, configured with the solar system's major planets as perturbing bodies and using the IAS15 integrator with an adaptive timestep \citep{rebound2012A&A...537A.128R,rein2015ias15}. Streams were modeled as packets of meteoroids, each originating from similar orbital elements but including small initial variations to simulate natural dispersion. The initial positions of the 100 particles in each packet were identical, but an initial uniform velocity variation between $0-5$\,m\,s$^{-1}$ was added in each x-y-z direction. The ejection point true anomaly on the orbit was also chosen from a uniformly random distribution. For each packet, the primary meteoroid defined the base orbit, with subsequent meteoroids perturbed slightly in velocity space to represent typical ejection velocities from a parent body. The semi-major axis and eccentricity space were mapped out with a grid layout, allowing the generation of 100 fictitious streams per inclination angle (0$^\circ$, 10$^\circ$, and 20$^\circ$). The simulation data was used to construct decoherence maps, illustrating the decoherence lifetimes as functions of semi-major axis and eccentricity for different inclinations.

The dynamical evolution of these particles was tracked over a simulation time of 5 million years, as the decoherence lifetimes of streams tend to be on the scale of 10$^{4}$-10$^{5}$\,years \citep{pauls2005decoherence}. The spread of orbital elements over time was monitored, focusing on the size of the largest coherent cluster within each packet. The DBSCAN (Density-Based Spatial Clustering of Applications with Noise) algorithm was employed to identify and track the sizes of these clusters as it has been previously demonstrated as an effective way to identify clusters in meteor datasets \citep{moorhead2016performance,sugar2017meteor}. The parameters for the DBSCAN algorithm included:
\begin{itemize}
    \item \textit{Epsilon ($\epsilon$):} The maximum distance between two points to be considered as neighbors. This value varied based on the specific D-function used and was based on the results of the NEO similarity significance analysis:
    \begin{itemize}
        \item $D_{SH}$: $\epsilon = 0.02$
        \item $D'$: $\epsilon = 0.03$
        \item $D_H$: $\epsilon = 0.03$
    \end{itemize}
    \item \textit{Minimum Points:} The minimum number of points required to form a core point in the cluster. This was set to 3, meaning each core point had to be associated with at least two other points.
\end{itemize}

The decoherence lifetime was defined as the time it took for the largest cluster to reduce to less than 5\% of its original size, indicating significant dispersion of the meteoroid stream.

\section{Results}\label{sec:results}
\subsection{KDE Fitting}
% describe briefly the fits and the bandwidths used 
In this study, we applied KDE to estimate the sporadic probability density functions (PDF) of the orbital distributions within our datasets. This approach is integral to assessing the statistical significance of orbital similarities, which will be explored in the next section. The KDE method was chosen because it effectively smooths the data to create a continuous representation of the underlying orbital distributions, possibly removing any small streams present in the data, while preserving the overarching features of each dataset.

\begin{figure*}
    \centering
    \includegraphics[width=\linewidth]{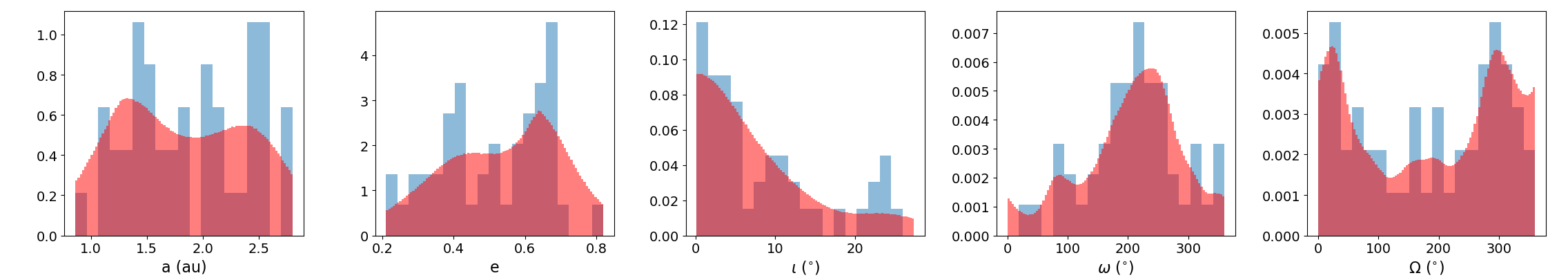}
    \caption{Orbital distribution of 50 recovered meteorite falls (blue) versus KDE estimated PDF (red) using a Gaussian kernel with a bandwidth of 0.5. This PDF, approximating the corresponding sporadic population, was used to estimate the degree of random association in the population via Monte Carlo simulations.}
    \label{fig:meteorite_kde}
\end{figure*}

\begin{figure*}
    \centering
    \includegraphics[width=\linewidth]{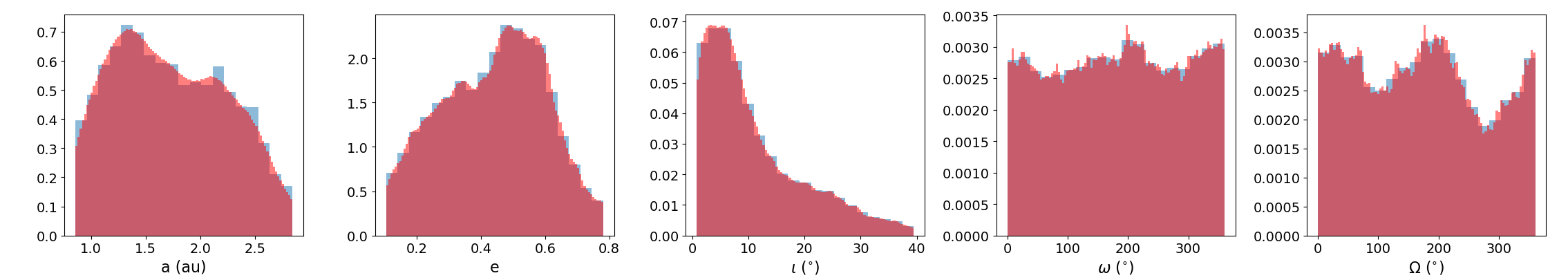}
    \caption{Orbital distribution of 35,012 NEOs (blue) versus KDE estimated PDF  (red) using a Gaussian kernel with a bandwidth of 0.05. This PDF, approximating the corresponding sporadic population, was used to estimate the degree of random association in the population via Monte Carlo simulations.}
    \label{fig:NEO_kde}
\end{figure*}

\begin{figure*}
    \centering
    \includegraphics[width=\linewidth]{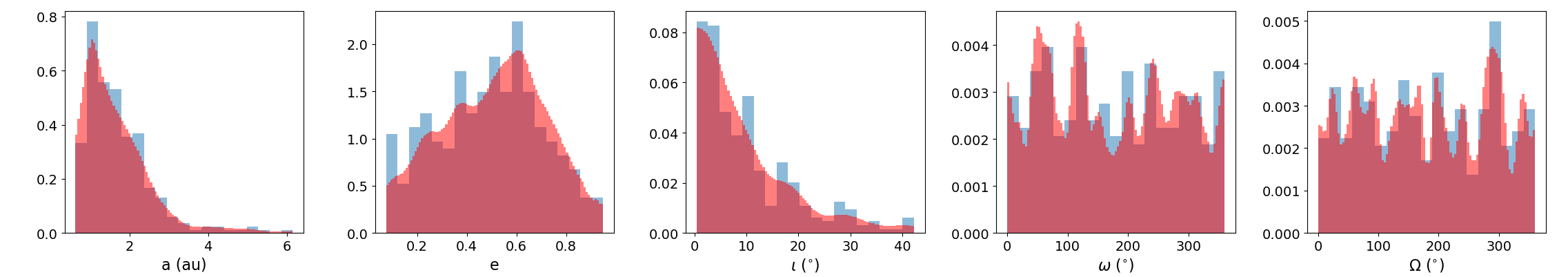}
    \caption{Orbital distribution of 310 US Government sensor impacts (blue) versus KDE estimated PDF (red) using a Gaussian kernel with a bandwidth of 0.15. This PDF, approximating the corresponding sporadic population, was used to estimate the degree of random association in the population via Monte Carlo simulations.}
    \label{fig:CNEOs_kde}
\end{figure*}

\begin{figure*}
    \centering
    \includegraphics[width=\linewidth]{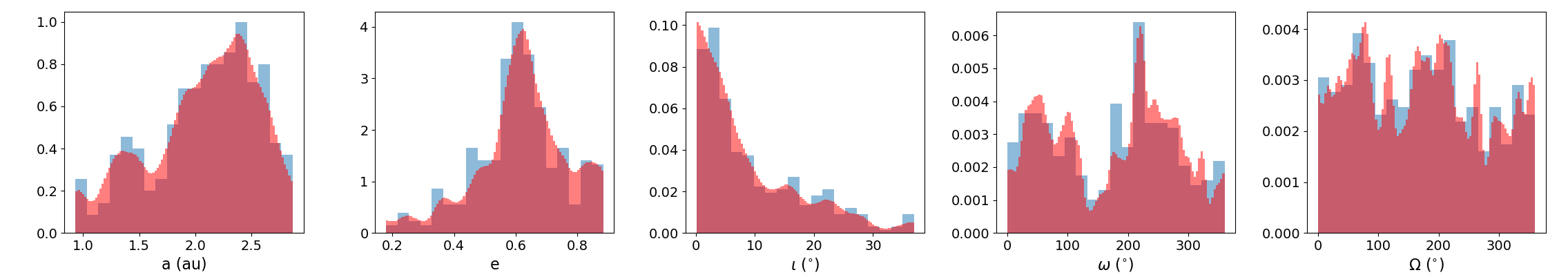}
    \caption{Orbital distribution of 350 probable 50\,g meteorite falls detected by GFO, FRIPON, or EFN sensors (blue) versus KDE estimated PDF  (red) using a Gaussian kernel with a bandwidth of 0.15 and ensuring any sample crossed the torus traced out by the Earth`s orbit. The meteorite falls were identified using $\alpha-\beta$ parameters as described in \citet{sansom2019determining} using a minimum end mass of 50\,g. This PDF, approximating the corresponding sporadic population, was used to estimate the degree of random association in the population via Monte Carlo simulations.}
    \label{fig:fall_kde}
\end{figure*}

The PDFs (seen as red in the figures) estimated by the KDE method are shown for the 50 meteorite falls (Fig.~\ref{fig:meteorite_kde}), the 35,012 NEOs (Fig.~\ref{fig:NEO_kde}), 350 potential 50\,g falls detected by GFO/FRIPON/EFN (Fig.~\ref{fig:fall_kde}), and the 310 USG Sensor detections (Fig.~\ref{fig:CNEOs_kde}). 

For the KDE fitting to be effective, it is crucial that the KDE accurately represents the general distribution of each dataset while smoothing out minor features such as small meteoroid streams or showers. This method is particularly suitable for datasets where showers constitute a minimal portion of the population. As demonstrated by \citet{shober2024generalizable}, the KDE approach may not be applicable when showers make up a significant fraction of the dataset, as their presence can distort the overall distribution. To address this, fireball datasets with significant shower components need to have these showers removed prior to analysis. This can be achieved using a dissimilarity function that does not directly rely on the orbital elements, such as the \citet{valsecchi1999meteoroid} function.

In our datasets used here, the fireball data included only events with a high likelihood of dropping meteorites, effectively eliminating all shower contributions. The degree of smoothing in the KDE is controlled by the bandwidth parameter. For datasets with limited data, the bandwidth is set higher to prevent overfitting and ensure that any small streams or showers present are adequately smoothed out. This approach allows us to detect and account for any minor clusters within the dataset, ensuring that the KDE fitting accurately reflects the broader orbital distribution.

We utilized the \texttt{KernelDensity} class from scikit-learn to execute random sampling based on a Gaussian KDE \citep{pedregosa2011scikit}. The sampling process begins by randomly choosing base points from the dataset involved in the KDE fit, allowing each data point an equal probability of being selected. If sample weights are specified, these probabilities are accordingly adjusted. Each base point chosen, denoted as $x_i$, is then perturbed by adding Gaussian noise, thereby generating a new sample point $x'_i$ calculated as $x'_i = x_i + N(0, h^2)$. Here, $N(0, h^2)$ represents Gaussian noise with zero mean and variance $h^2$, where $h$ is the KDE bandwidth parameter. This parameter is crucial as it modulates the spread of the generated samples around their corresponding base points, thereby influencing the smoothness of the resultant density estimation. Each iteration of this procedure ensures that the sampled points closely mirror the density profile estimated by the KDE, using the Gaussian kernel’s properties to facilitate tasks such as Monte Carlo simulations and the generation of synthetic datasets. Detailed implementation specifics of scikit-learn's \textit{KernelDensity} class can be accessed through its GitHub repository\footnote{\url{https://github.com/scikit-learn/scikit-learn/blob/9e38cd00d/sklearn/neighbors/_kde.py#L35}}. This methodology of using KDE to estimate the random association rate between orbits was first employed by \citet{shober2024generalizable}. 

\begin{figure*}
    \centering
    \includegraphics[width=\linewidth]{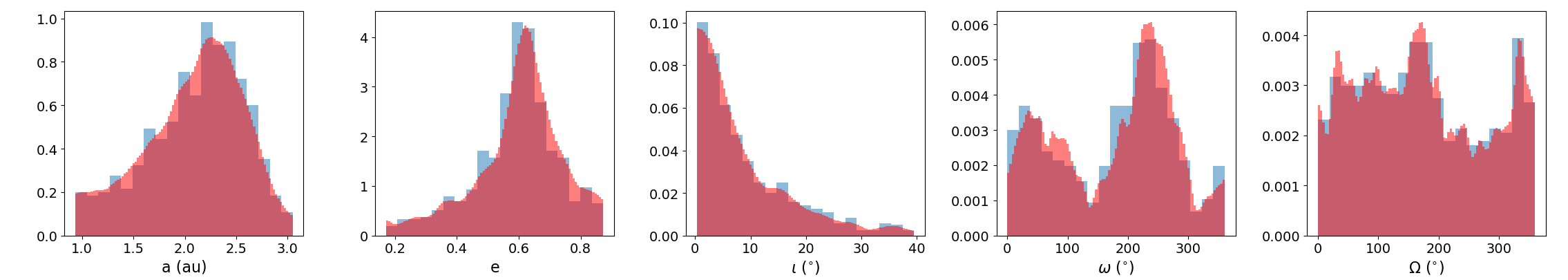}
    \caption{Orbital distribution of 616 possible meteorite falls detected by GFO, FRIPON, or EFN sensors (blue) versus KDE estimated PDF (red) using a Gaussian kernel with a bandwidth of 0.15 and ensuring any sample crossed the torus traced out by the Earth`s orbit. The meteorite falls were identified using $\alpha-\beta$ parameters as described in \citet{sansom2019determining} using a minimum end mass of 1\,g. This PDF, approximating the corresponding sporadic population, was used to estimate the degree of random association in the population via Monte Carlo simulations.}
    \label{fig:616_1g_kde}
\end{figure*}

This random sampling from the estimated PDFs produced through KDE notably differs from previous studies that explored the chance similarity of orbits. Notably, the study of \citet{jopek2017probability} primarily recommended the use of cumulative probability distribution (CPD) inversion to generate random samples by mapping uniform random variables to quantiles of the target distribution using the inverse of its cumulative distribution function (CDF). This method ensures that the generated samples accurately reflect the overall distribution's quantiles. In contrast, KDE smooths observed data to estimate the PDF, providing a continuous approximation that highlights the density's shape and local features while preserving parameter correlations. If some of the local features are real clusters and the CPD inversion includes them as part of the PDF to estimate random associations, the statistical significance of the cluster will be underestimated as the cluster was assumed to be a feature of the sporadic background. Meanwhile, the KDE's ability to smooth out the possible presence of small clusters in the dataset when estimating the PDF removes this underestimation. With that in mind, choosing a bandwidth that is too large for the KDE and over-smoothing the PDF could equally result in an overestimation of the significance of a cluster. Thus, regardless of the method chosen, one must be aware of these inherent setbacks/strengths. 

In \citet{pauls2005decoherence}, they constructed two randomized orbital distributions by using the debiased NEO orbital distribution published by \citet{bottke2002debiased} and from a dataset of 481 fireballs. They randomly selected (a, e, i) triplets from these distributions and used a pseudo-random number to determine if the selected triplet would impact Earth (only for NEO model). They then selected random longitudes of the ascending node uniformly from 0 to 360$^{\circ}$ and fixed the argument of perihelion such that either the ascending or descending node would be within the torus swept out by the Earth. This allowed them to construct databases of orbits with randomized angular elements but realistic semi-major axis, eccentricity, and inclination values. Albeit, this method also could suffer from the same slight underestimation of small clusters as discussed possible for the method of \citet{jopek2017probability}. 

As seen in Fig.~\ref{fig:meteorite_kde}, due to the low statistics (only 50 orbits) of the meteorite fall with orbits dataset, the KDE's smoothing can reduce the small dataset's considerable variation and produce a more reasonable continuous distribution to draw from. The larger bandwidth of 0.5 used for the meteorite falls seems to produce an adequate distribution and is larger due to the small size of this specific dataset. The known NEO distribution comprises 35,012 telescopically observed objects (Fig.~\ref{fig:NEO_kde}). Meanwhile, the impact database detected by USG Sensors has 310 events (Fig.~\ref{fig:CNEOs_kde}) with velocity information available. Based on thousands of observations from the FRIPON, EFN, and GFO networks, 615 possible $>$\,1\,g meteorite-dropping fireballs were detected (Fig.~\ref{fig:616_1g_kde}), of which 315 fireballs are possible $>$\,50\,g meteorite-droppers (Fig.~\ref{fig:fall_kde}). 

\subsection{Statistical Significance of Orbital Similarity}
\subsubsection{Fireball Clustering}
% start with quick summary of Pauls & Gladman (2005) significance results
In \citet{pauls2005decoherence}, they found that despite previous claims, the probability that a chance association as similar as Příbram and Neuschwanstein ($D'\sim$\,0.009) was expected to occur $\cong$70\% of time for a fireball dataset of 481 orbits. Even using another fireball model based upon a model NEA distribution \citet{bottke2002debiased} still gave a chance occurrence of a $D'$ = 0.009 to be 10\%, still not low enough to reject the null hypothesis. Furthermore, the other streams suggested by \citet{halliday1990evidence} were not identified as significantly lower than expected from simple random associations. 

\begin{figure}[]
    \centering
    \begin{subfigure}[b]{0.5\textwidth}
        \includegraphics[width=\textwidth]{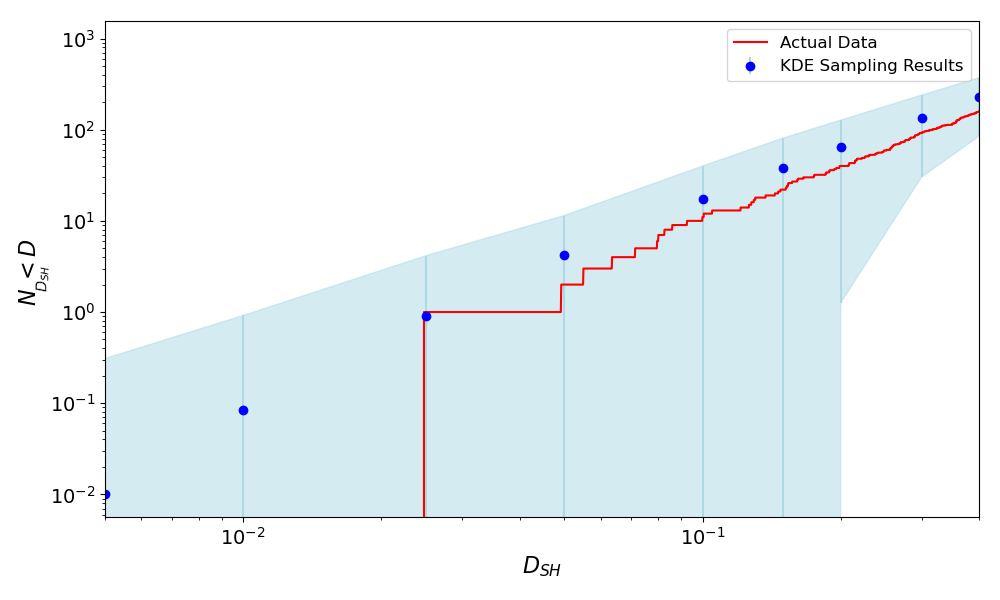}
        % \caption{$\iota$=0$^{\circ}$}
        \label{fig:d_sh_meteorite_falls}
    \end{subfigure}
    
    \begin{subfigure}[b]{0.5\textwidth}
        \includegraphics[width=\textwidth]{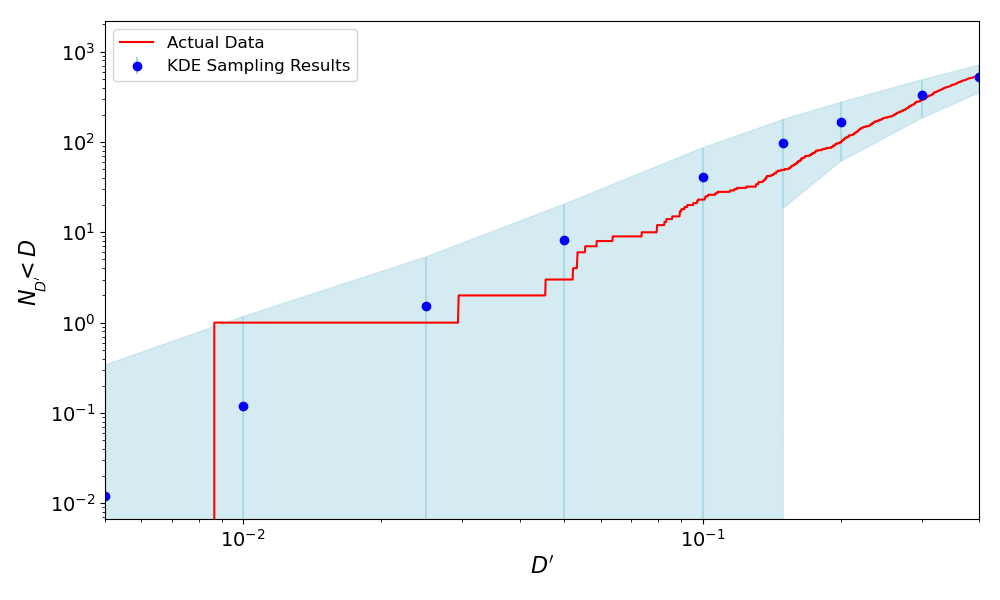}
        % \caption{$\iota$=10$^{\circ}$}
        \label{fig:d_prime_meteorite_falls}
    \end{subfigure}

    \begin{subfigure}[b]{0.5\textwidth}
        \includegraphics[width=\textwidth]{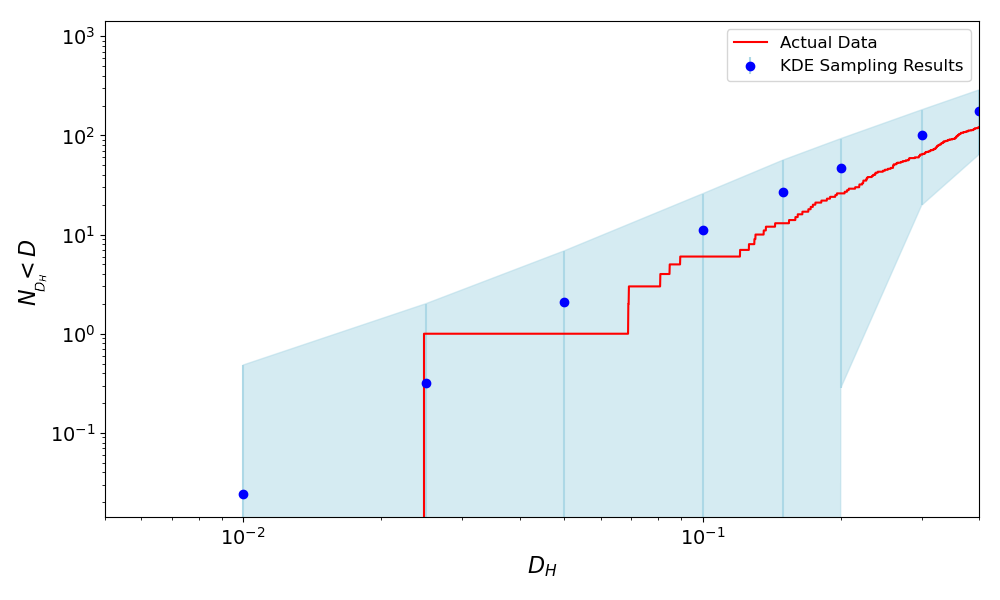}
        % \caption{$\iota$=20$^{\circ}$}
        \label{fig:d_h_meteorite_falls}
    \end{subfigure}
    \caption{The number of meteorite-meteorite pairs, N$_{\text{D}}$, with D values below a limiting threshold (red line) compared to the predicted number based on random associations (blue circles with error bars) for $D_{SH}$, $D'$, and $D_{H}$. The blue circles represent the mean number of pairs from 500 iterations of KDE sampling using a bandwidth of 0.1 based on 350 possible/likely 50\,g meteorite falls observed by the GFO, FRIPON, and EFN networks. The blue-shaded area is the 3\,$\sigma$ range possible from random associations. The actual number of low-D value pairs is consistent with the predicted number based on random associations.}
    \label{fig:false_positive_meteorites_meteorites}
\end{figure}

In this study, we have repeated this statistical analysis, using a similar KDE-based method to estimate the false positive rate as done in \citet{shober2024comparing}. Our model of the fireball population was fitted to 350 identified potential meteorite-dropping fireballs observed by the EFN, GFO, and FRIPON. This subset was selected through a combination of several metrics and methods, including the $\alpha-\beta$ methodology of \citet{sansom2019determining}, the fragmentation/ablation models used in \citet{borovivcka2022_one}, shower identification and removal \citep{valsecchi1999meteoroid}, and sporadic cometary component removal \citep{tancredi2014criterion,shober2021main,shober2024comparing}. By randomly drawing 481 orbits from the PDF generated from this meteorite dropper distribution, we find that there is a $\sim$3.1\% chance of a random association. Despite this estimate being lower than that of \citet{pauls2005decoherence}, in the nearly 20 years since their study, there have been no more associations of that degree within the fall dataset. Thus, the likelihood of a chance association is higher than this, considering the many new and ongoing global fireball networks; the number of observed falls is very likely to be over a thousand or more. More importantly, we find that using our KDE false positive method, the meteorite-meteorite similarities using all three D-functions are within a 3-$\sigma$ range expected for random associations (Fig.~\ref{fig:false_positive_meteorites_meteorites}). The Příbram-Neuschwanstein pair only seems less likely for the $D'$ discriminant, while the other two metrics produce distributions that are very consistent with a spurious association. 

The orbital distribution for the 50 recovered falls (Fig.~\ref{fig:meteorite_kde}), however, seems to significantly deviate from the distribution of predicted meteorite falls observed by fireball networks (Fig.~\ref{fig:616_1g_kde},\ref{fig:fall_kde}). Thus, this distribution either (1) does not estimate well the population, or (2) searching/finders bias is significantly modifying the orbital distribution of recovered falls. When only using the 50 recovered fall data points to reconstruct the fall distribution, the statistical significance increases, at least when using a bandwidth of 0.5. However, using this KDE method on such a small dataset, while possible, is neither ideal nor trivial in selecting the bandwidth parameter well. It is much easier to accidentally over- or under-smooth the true distribution, leading to less precise results and possible false detections or false non-detections of streams. Thus, a larger dataset of 350 possible $>$\,50\,g meteorite falls was used to estimate the fall population more precisely (Fig.~\ref{fig:false_positive_meteorites_meteorites}), and the resulting correspondence between the observations and random association rate is considered more robust. Moreover, another critical factor is that there is a more general similarity amongst all possible pairs of the 50 recovered meteorites, as supported by the presence of a significantly shallower cumulative D-value slope, supporting the hypothesis that a search/recovery bias is affecting this recovered fall sample. 

\begin{figure}[]
    \centering
    \begin{subfigure}[b]{0.5\textwidth}
        \includegraphics[width=\textwidth]{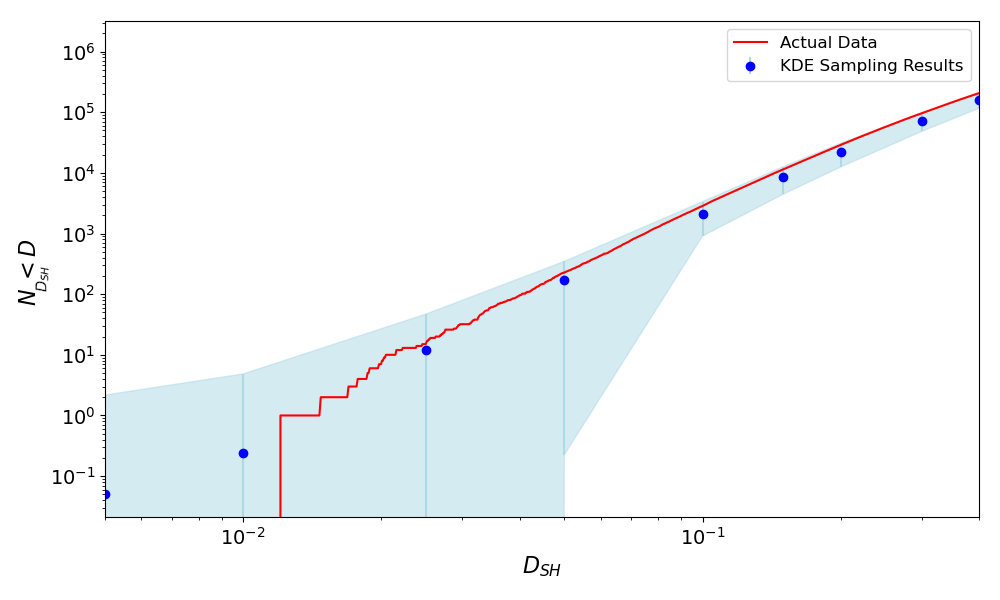}
        % \caption{$\iota$=0$^{\circ}$}
        \label{fig:d_sh_meteorite_neo}
    \end{subfigure}
    
    \begin{subfigure}[b]{0.5\textwidth}
        \includegraphics[width=\textwidth]{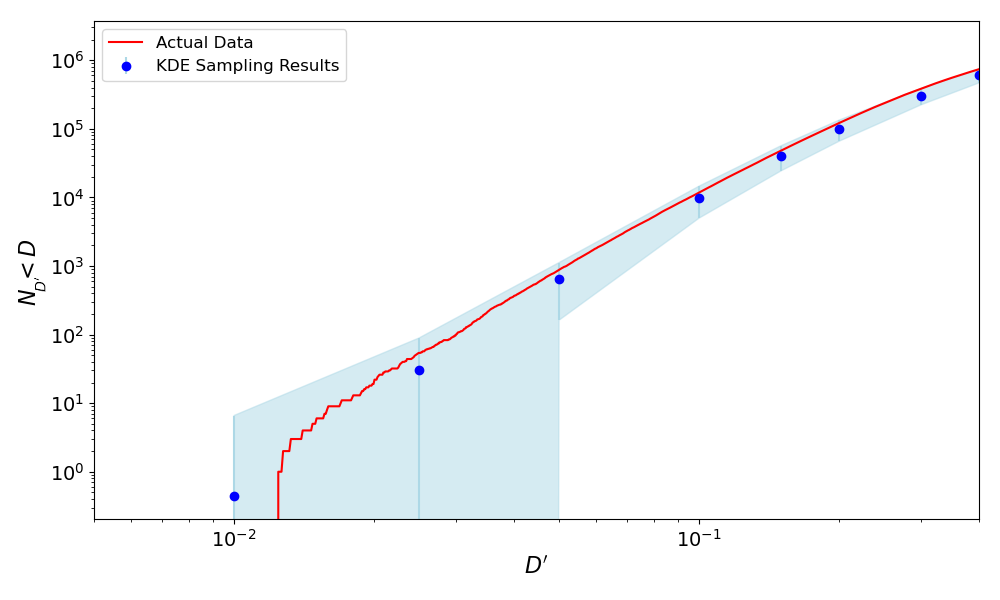}
        % \caption{$\iota$=10$^{\circ}$}
        \label{fig:d_prime_meteorite_neo}
    \end{subfigure}

    \begin{subfigure}[b]{0.5\textwidth}
        \includegraphics[width=\textwidth]{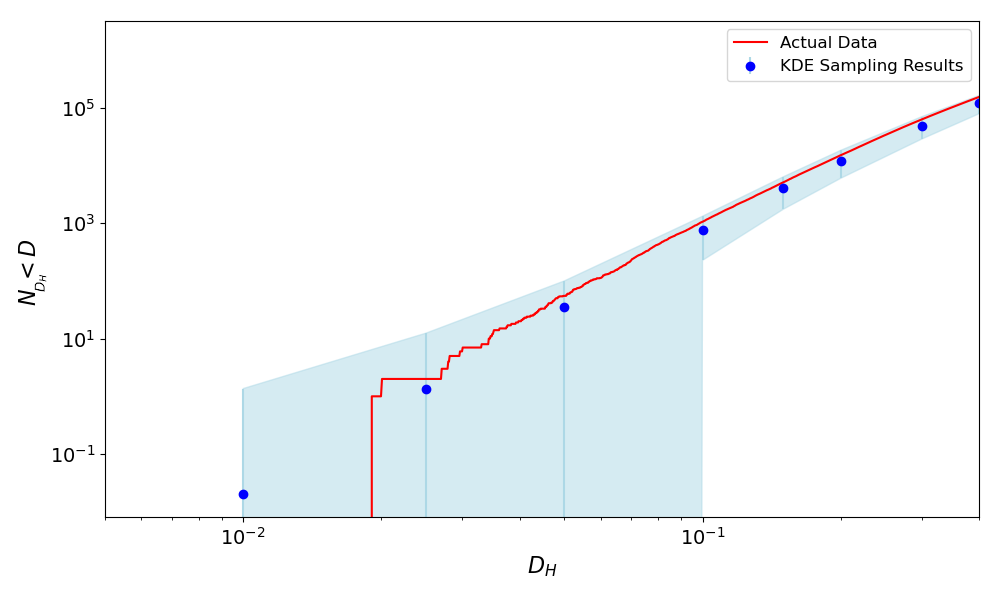}
        % \caption{$\iota$=20$^{\circ}$}
        \label{fig:d_h_meteorite_neo}
    \end{subfigure}
    \caption{The number of meteorite-NEO pairs, N$_{\text{D}}$, with D values below a limiting threshold (red line) compared to the predicted number based on random associations (blue circles with error bars) for $D_{SH}$, $D'$, and $D_{H}$. The blue circles represent the mean number of pairs identified between 50 random samples of ``meteorite falls'' with the observed 35,012 NEOs from 500 iterations of KDE sampling using a bandwidth of 0.15 for the 350 possible 50\,g meteorite falls (Fig.~\ref{fig:fall_kde}). The blue-shaded area is within three standard deviations for random samples of 50 (the number of meteorites with orbits). The actual number of low-D value pairs correctly follows the predicted number based on random associations, indicating no evidence supporting the presence of streams in the recovered meteorite fall dataset. Two asteroids were removed as they correspond to two recovered meteorite falls, Motopi Pan and Almahata Sitta, observed telescopically before impact.}
    \label{fig:false_positive_meteorites_NEOs}
\end{figure}

\begin{figure}[!htbp]
    \centering
    \begin{subfigure}[b]{0.5\textwidth}
        \includegraphics[width=\textwidth]{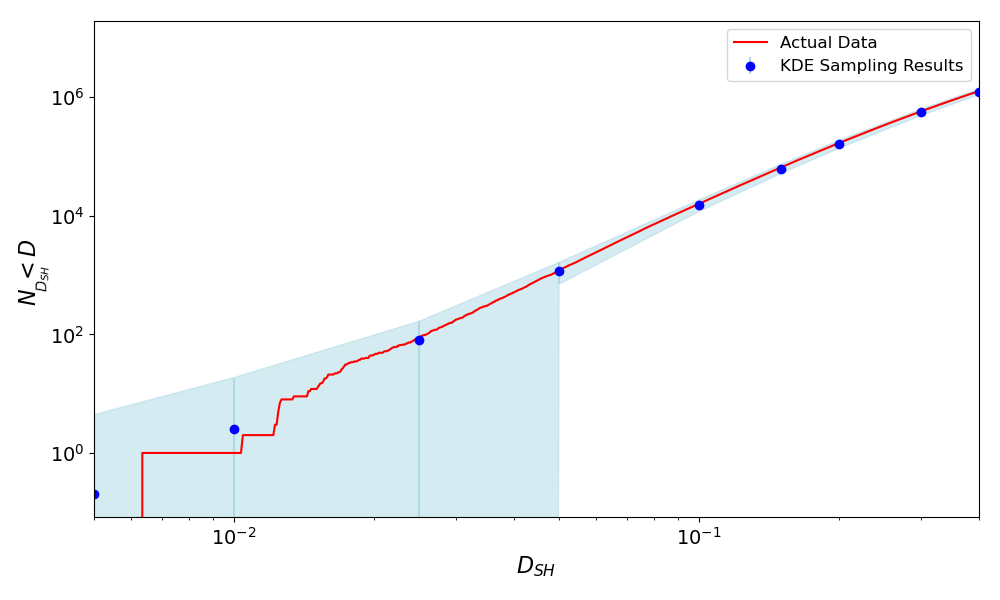}
        % \caption{$\iota$=0$^{\circ}$}
        \label{fig:d_sh_cneos_neo}
    \end{subfigure}
    
    \begin{subfigure}[b]{0.5\textwidth}
        \includegraphics[width=\textwidth]{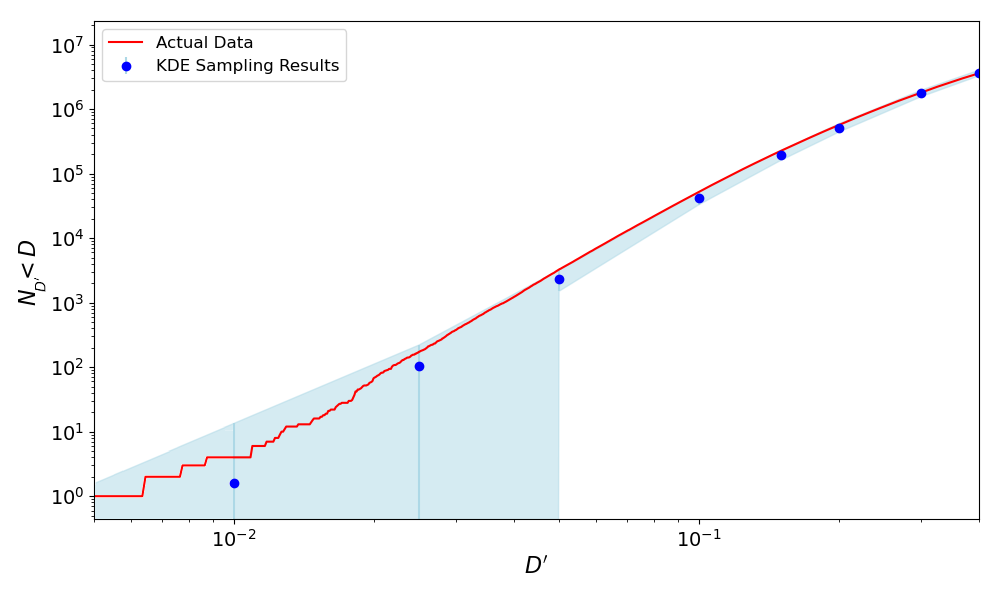}
        % \caption{$\iota$=10$^{\circ}$}
        \label{fig:d_prime_cneos_neo}
    \end{subfigure}

    \begin{subfigure}[b]{0.5\textwidth}
        \includegraphics[width=\textwidth]{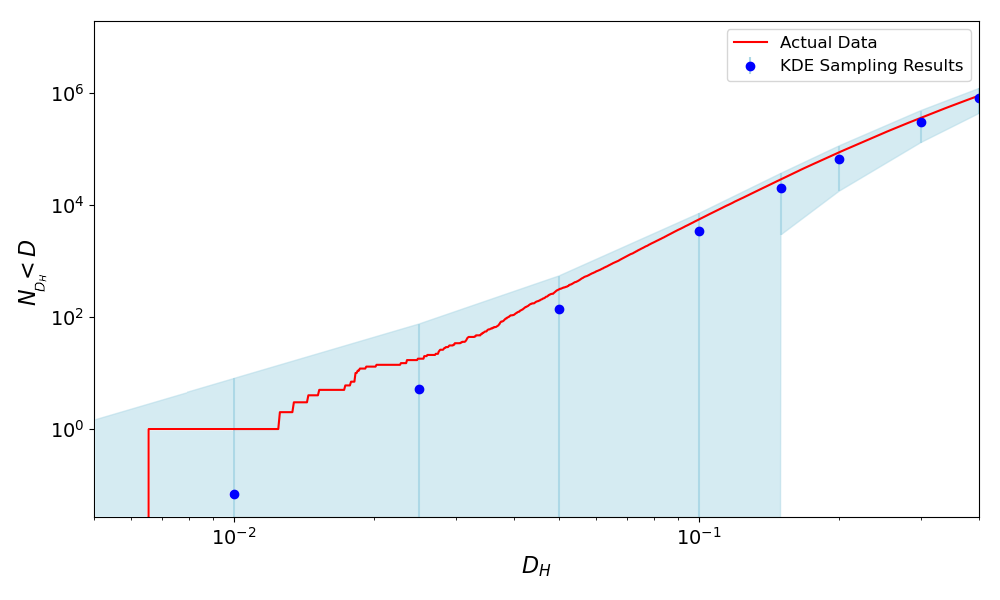}
        % \caption{$\iota$=20$^{\circ}$}
        \label{fig:d_h_cneos_neo}
    \end{subfigure}
    \caption{The number of CNEOS USG Sensors impact - NEO pairs, N$_{\text{D}}$, with D values below a limiting threshold (red line) compared to the predicted number based on random associations (blue circles with error bars) for $D_{SH}$, $D'$, and $D_{H}$. The blue circles represent the mean number of pairs identified between 310 random samples drawn from the USG impactors PDF (Fig.~\ref{fig:CNEOs_kde}) with the observed 35,012 NEOs from 500 iterations of KDE sampling. The blue-shaded indicates a zone encompassing three standard deviations. The number of low-D value pairs corresponds very well to the predicted number based on random associations using the KDE distribution shown in Figs.~\ref{fig:NEO_kde}, indicating no substantial evidence supporting streams in the dataset.}
    \label{fig:false_positive_CNEOS_NEOs}
\end{figure}

\begin{figure}[!htbp]
    \centering
    \begin{subfigure}[b]{0.5\textwidth}
        \includegraphics[width=\textwidth]{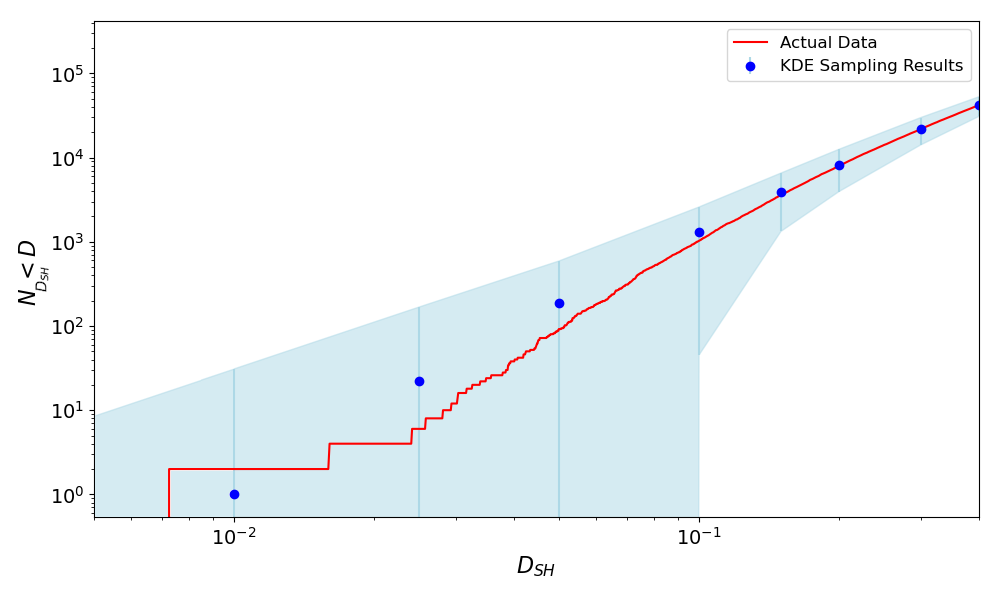}
        % \caption{$\iota$=0$^{\circ}$}
        \label{fig:d_sh_cneos_neo}
    \end{subfigure}
    
    \begin{subfigure}[b]{0.5\textwidth}
        \includegraphics[width=\textwidth]{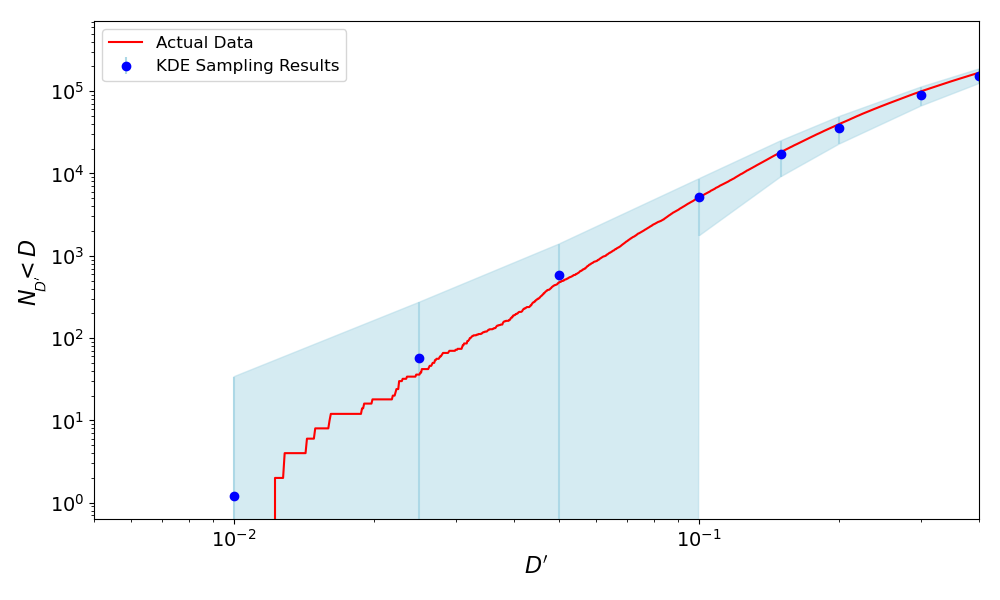}
        % \caption{$\iota$=10$^{\circ}$}
        \label{fig:d_prime_cneos_neo}
    \end{subfigure}

    \begin{subfigure}[b]{0.5\textwidth}
        \includegraphics[width=\textwidth]{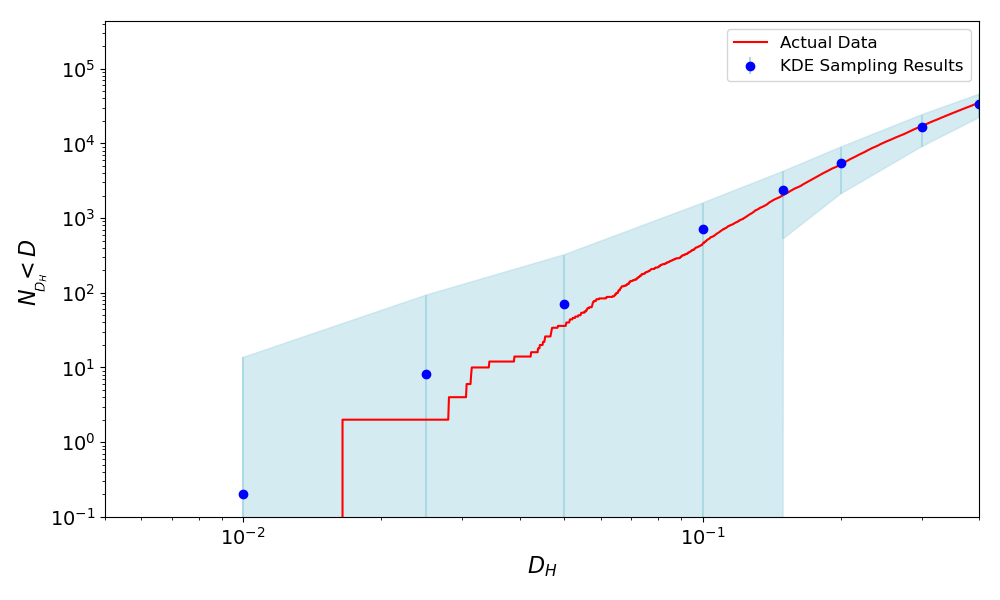}
        % \caption{$\iota$=20$^{\circ}$}
        \label{fig:d_h_cneos_neo}
    \end{subfigure}
    \caption{The number of unique pairs, N$_{\text{D}}$, within the 616 possible 1\,g final mass fireballs observed by the GFO/EFN/FRIPON with D values below a limiting threshold (red line) compared to the predicted number based on random associations (blue circles with error bars) for $D_{SH}$, $D'$, and $D_{H}$. The blue circles represent the mean number of pairs identified between two 616 random samples drawn from the possible 1\,g fall PDF (Fig.~\ref{fig:616_1g_kde}) from 500 iterations of KDE sampling. The blue-shaded indicates a zone encompassing three standard deviations. The number of low-D value pairs corresponds very well to the predicted number based on random associations, indicating no substantial evidence supporting streams in the fireball dataset.}
    \label{fig:false_positive_616_fireballs}
\end{figure}

\begin{figure*}[]
    \centering
    \begin{subfigure}[b]{0.49\textwidth}
        \includegraphics[width=\textwidth]{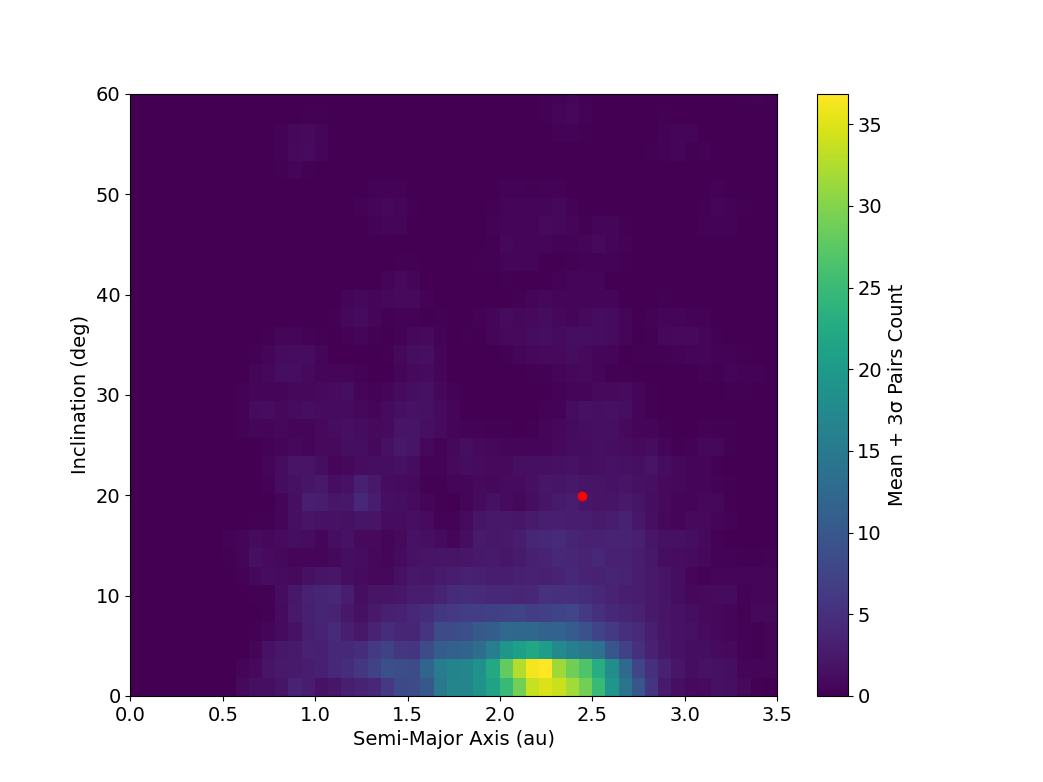}
        \label{fig:ai_1g}
    \end{subfigure}%
    % \hspace{0.01\textwidth}
    \begin{subfigure}[b]{0.49\textwidth}
        \includegraphics[width=\textwidth]{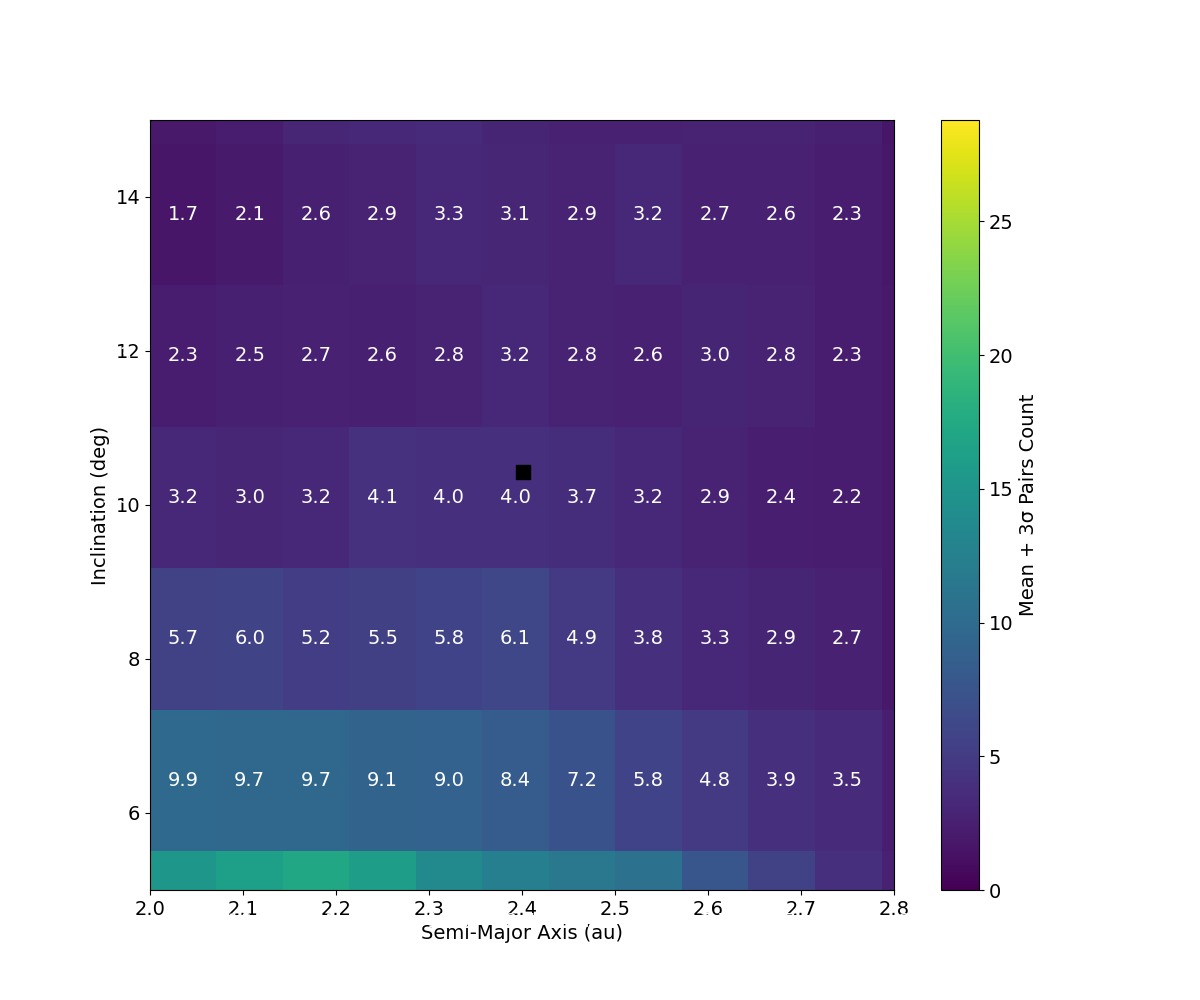}
        \label{fig:ai_pribram}
    \end{subfigure}%
    \caption{Maximum false positive heat map for bins of semi-major axis (au) by inclination (deg). The heat map coloration represents the mean number of pairs plus 3$\sigma$ that have D$_{H}<0.1$. The sporadic orbital distribution was estimated using the 616 possible 1\,g fall observations PDF (Fig.~\ref{fig:616_1g_kde}). The means and standard deviations were calculated through 500 iterations of KDE sampling. Subplot a examines the maximum number of expected random associations within a population of 616 fireballs, only one (red) in the observations has a measured D$_{H}<0.1$, but it is not statistically significant. Subplot b re-examines the significance of the Příbram/Neuschwanstein pair (black square); however, the pair is still not statistically significant within a fireball population of 481 orbits.}
    \label{fig:ai_significance}
\end{figure*}

% Explain the false positive figure layout and the results 
As explained in further detail in section~\ref{sec:method}, our Monte Carlo KDE sampling method enabled us to fully explore the orbital similarity within the fireball, fall, and NEO datasets. The cumulative D-value distribution figures produced from this method demonstrate how all unique pair similarities within a single population or between two populations compare to those expected from random associations. The blue shaded areas in Figures~\ref{fig:false_positive_meteorites_meteorites}-\ref{fig:false_positive_NEOS_NEOs} indicate the 3-$\sigma$ regions given by our model consistent with the range of values produced by random associations. Meanwhile, the red lines indicate the actual observed cumulative D-value distribution. 

First, we compared the similarity significance between the 50 meteorite falls and the known NEO population. Nearly every study of recovered meteorite falls with precisely measured orbits has tried to find NEOs that are in similar orbit space. Despite problems dealing with CRE ages and decoherence lifetimes, which we will address in section~\ref{sec:discuss}, are any of these associations statistically significant? 

These results are illustrated in Fig.\ref{fig:false_positive_meteorites_NEOs}. The Monte Carlo simulation aimed to estimate the number of random associations expected for each of the three D-functions used here ($D_{SH}$, $D'$, $D_H$) using a dataset of 350 possible 50\,g meteorite falls observed by the GFO, EFN, and FRIPON to estimate the orbital distribution. As depicted in the figure, the expected number of random associations closely matches the actual data, indicating that the orbital similarity between the meteorite databases (50 meteorite falls) and the NEOs is entirely consistent with random chance associations. This finding contradicts several studies that have used low D-values with NEOs to suggest associations between meteorites and near-Earth objects. Similar results were also found when the meteorite fall PDF was estimated using the orbits from the 50 recovered meteorites population (Fig.~\ref{fig:meteorite_kde}), and the 616 possible 1\,g falls (Fig.~\ref{fig:616_1g_kde}). No matter the fireball subset used, we found no statistically significant clustering between the fall and NEO populations. Our results demonstrate that the associations are consistent with random chance throughout the D-value ranges. One interesting aspect, however, is the noticeable overestimation of the KDE at larger D-values. We also tested the random association likelihood drawing from a KDE fit to a very smoothed (bandwidth=0.5; Fig.~\ref{fig:meteorite_kde}) distribution of the 50 recovered meteorite falls, and this also produced cumulative D-value distribution well within the 3$\sigma$ region of random associations. 

D-values have also been used to identify the parent or associated asteroids from the NEO population with USG sensor-detected impacts. Thus, we also compared this population with the NEO population to estimate the number of random associations as a function of the D-value. Fig.~\ref{fig:false_positive_CNEOS_NEOs} shows that the data is consistent with random associations for all three metrics. This analysis indicates no evidence of statistically significant similarity between the CNEOS database and the NEO population. 

% add 616 1g falls statistical significance 
The CNEOS bolide database, which uses USG sensor observations, is widely known now to have substantial and seemingly unpredictable orbital precision, and no formal uncertainties are associated \citep{devillepoix2019observation}. On the other hand, for fireballs observed by dedicated photographic networks, where the orbits and their corresponding uncertainties are understood, can we find statistically significant clustering? Using a lower final mass limit of $\sim$1\,g to augment the size of the dataset, we collected 616 fireballs from GFO/EFN/FRIPON observations (Fig~\ref{fig:616_1g_kde}). Despite the increased precision of this data, as seen in Fig.~\ref{fig:false_positive_616_fireballs}, the number of associations is also completely in correspondence with that expected from random associations. For the moment, there seems to be no evidence based on orbital similarity that streams exist in the centimeter- to meter-sized populations that are impacting the Earth's atmosphere.  

% talk about statistical significance as a function of a-i heat map 
% talk about intuition that a low-D for an object near ecliptic is less significant, whereas the same value for high-inclination is more.... we now characterized that
In the analysis of fireball and meteorite recoveries, there is often a logical assumption that low D-values are less significant for bodies with inclinations near the ecliptic due to the high density of such orbits. In contrast, similar D-values for higher inclination orbits are considered more meaningful. In this study, we have quantified how the statistical significance of these D-values varies as a function of the orbit as well. Thus far, we have only examined whether the total number of orbital pairs is below a certain threshold. However, we must also consider the probability of getting an orbital pair below some D-threshold within the context of the orbital space of the pairs to avoid the missed identification of small clusters in low-probability orbits (e.g., high inclinations). It has also been shown recently how theoretical NEO families disperse more slowly when they are formed at higher inclinations and further from the nearest planet \citep{humpage2024numerical}. Thus, any clusters on these orbits should persist longer, and the examination of the statistical significance as a function of the orbit cannot be overlooked.

Nevertheless, as seen in Fig.~\ref{fig:ai_significance}, we find that even when the statistical significance is tested as a function of the orbit, we still find no pairs that exceed a 3$\sigma$ limit. To construct Fig.~\ref{fig:ai_significance}, we employed the same methodology as used in creating Fig.~\ref{fig:false_positive_616_fireballs} using the data of the 616 GFO/EFN/FRIPON fireballs. However, we also divided the orbital space into bins, and for each bin, we tracked the number of pairs with D$_{H}<0.1$. The heat map coloration in Fig.~\ref{fig:ai_significance} then shows the mean number of such pairs over all the Monte Carlo trials, plus three standard deviations. This approach allowed us to visualize where, within the orbital space, the statistical significance might deviate from random expectations. We also tested this mapping for D$_{H}<0.05$ and D$_{H}<0.1$ and logically encountered the same results, i.e., no statistically significant groupings.

Our findings reveal that none of the fireball pairs within this dataset exceed the 3-sigma limit, indicating no statistically significant clustering, even when accounting for the specific orbital contexts. We find only one pair (red point) amongst the 616 possible 1\,g falls that satisfies the D$_{H}<0.1$ used to make the heat map in Fig.~\ref{fig:ai_significance}, but it does not surpass a 3$\sigma$ significance threshold. Additionally, we re-considered the pair of Příbram/Neuschwanstein (black square) in Fig.~\ref{fig:ai_significance}. However, this too also failed to meet a 3$\sigma$ significance threshold when considering a dataset of 481 orbits, as done in \citet{pauls2005decoherence}. A similar analysis of the CNEOS dataset identified one impact event above the 3-sigma threshold at a 10$^{\circ}$ inclination. However, this result becomes inconsequential when considering the inherent unconstrained orbital uncertainties in the CNEOS dataset. Another similar claim was made for the Chelyabinsk impactor and NEA 1999 NC43; however, based on compositional and spectral information, this link was considered unlikely \citep{reddy2015link}. Beyond this non-correspondence, the statistical significance test used previously has some pretty large implicit assumptions that we would like to discuss. The statistical significance was determined by calculating the probability of drawing a randomly selected asteroid that is as similar to the measured Chelyabinsk orbit, only comparing to the asteroids with larger diameters than 1999 NC43 \citep{borovivcka2013trajectory}. To robustly reject this null hypothesis, one must assume that Chelyabinsk is not unique and is another random draw from the same source population of all other observed falls. This is the same problem as the ``birthday paradox'' discussed in \citet{pauls2005decoherence}. This veridical paradox states that within a group of 23 people or more, there is a greater than 50\% chance of two people sharing a birthday. This is not a true paradox, but rather just an unintuitive result, as most people approach this problem by trying to calculate the probability that they will personally share a birthday with someone in the group. This is the same logic used when only testing the probability of similarity with a specific orbit instead of the probability of a random association between any of the combinations of random draws of the concerned population(s). Additionally, the limitation of the asteroid population to only bodies greater than the size of 1999 NC43 significantly increases the significance of the association without adequate justification of the orbital uniqueness of this larger asteroid subset. In this study, we have redone this significance test using our KDE-based methodology. From this we find that when using similar assumptions \citep{borovivcka2013trajectory,reddy2015link}, similar significance values are found. However, considering that there were 20 recovered meteorites at the time of the Chelyabinsk impact and assuming this sample is being drawn from the same source population, the probability of a random association is still more than 3$\sigma$. However, if the size of the asteroid population includes smaller bodies, the orbital similarity is of 1999 NC43, and the Chelyabinsk impactor becomes insignificant. Thus, when calculating the statistical significance of two bodies, it is important to consider the assumptions carefully. 

Thus, by quantifying the relationship between D-value significance and orbital parameters, we find no robust evidence of statistically significant clustering. This further supports the conclusion that observed similarities are consistent with random associations rather than true physical streams in impact datasets. 

% talk about NEO-NEO pairs and cite Jopek, QQn+Granvik, etc. 
\subsubsection{NEO Clustering}
\begin{figure}[!htbp]
    \centering
    \begin{subfigure}[b]{0.5\textwidth}
        \includegraphics[width=\textwidth]{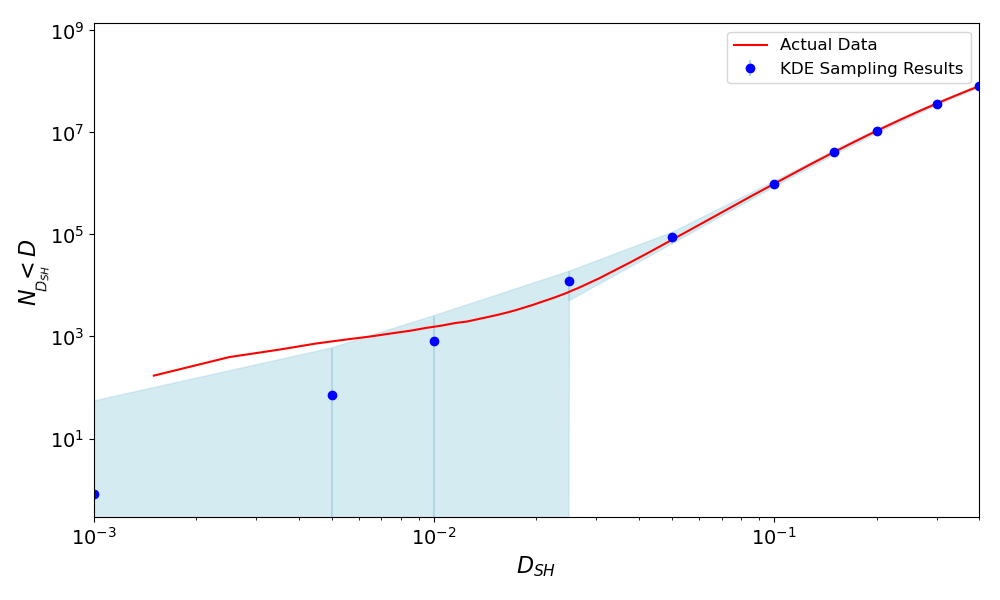}
        % \caption{$\iota$=0$^{\circ}$}
        \label{fig:d_sh_neos_neo}
    \end{subfigure}
    
    \begin{subfigure}[b]{0.5\textwidth}
        \includegraphics[width=\textwidth]{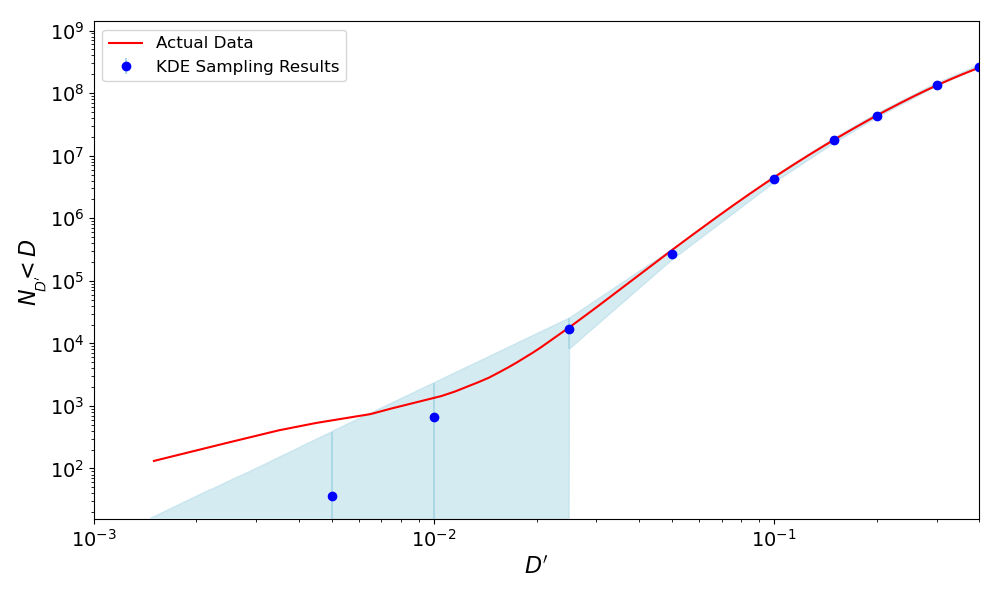}
        % \caption{$\iota$=10$^{\circ}$}
        \label{fig:d_prime_neo_neo}
    \end{subfigure}

    \begin{subfigure}[b]{0.5\textwidth}
        \includegraphics[width=\textwidth]{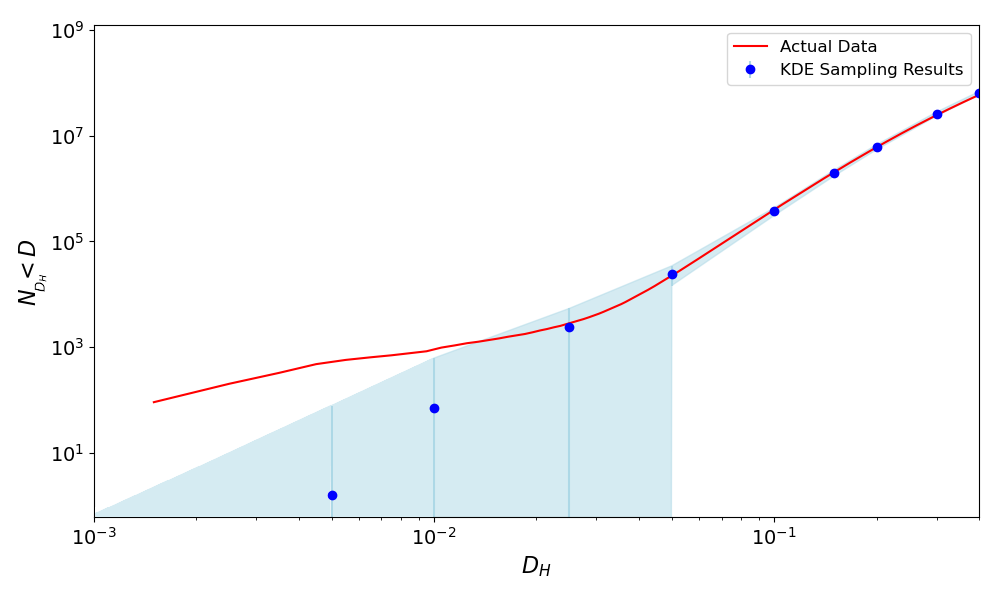}
        % \caption{$\iota$=20$^{\circ}$}
        \label{fig:d_h_neo_neo}
    \end{subfigure}
    \caption{The number of NEO - NEO pairs with D values below a limiting threshold (red line) compared to the predicted number based on random associations (blue circles with error bars) for $D_{SH}$, $D'$, and $D_{H}$. The blue circles represent the mean number of pairs from 500 iterations of KDE sampling using a bandwidth of 0.15, with a shaded blue region indicating a zone encompassing three standard deviations. The number of low-D value pairs corresponds very well to the predicted number based on random associations for large values; however, at very low D-values, there appear to be tens of objects with statistically significant similarities. The observational biases of the dataset do not predict the elbow in the cumulative D-value plots.}
    \label{fig:false_positive_NEOS_NEOs}
\end{figure}

Several investigations have also explored the presence of NEO families \citep{fu2005identifying,schunova2012searching,jopek2020orbital}. Using our KDE-based methodology, we have re-examined the statistical significance of orbital similarity among all unique pairs within the 35,012 NEOs from the NASA HORIZONS database. Our analysis revealed a remarkable consistency in the distribution of large D-values across all three orbital similarity discriminants (D$_{SH}$, D$_H$, and D'). This is clearly illustrated in Figure~\ref{fig:false_positive_NEOS_NEOs}, where the observed cumulative D-value distribution shows a distinct deviation from the random association model, particularly around D$_H$ = 0.03. This deviation, or "kink," which is present across all similarity distributions, indicates the presence of statistically significant groupings within the NEO population -- a feature that contrasts sharply with the expected results from random associations. This kink is also not completely explained due to the observational bias of telescopic surveys, as that would have been captured in the blue 3-$\sigma$ region in Fig.~\ref{fig:false_positive_NEOS_NEOs}. The statistical significance of these groupings begins to emerge at D-values around 10$^{-2}$, consistent across the three discriminants, although with some variation. Previous studies, such as \citet{schunova2012searching}, found no consistent evidence for NEO families, while more recent work by \citet{jopek2020orbital} identified 15 potential groupings with over 900 objects. Our results corroborate these more recent findings that statistically significant clustering is occurring, albeit we find that the clusters identified in \citet{jopek2020orbital} are much more inclusive than we find here. 

\begin{figure*}[]
    \centering
    \begin{subfigure}[b]{0.49\textwidth}
        \includegraphics[width=\textwidth]{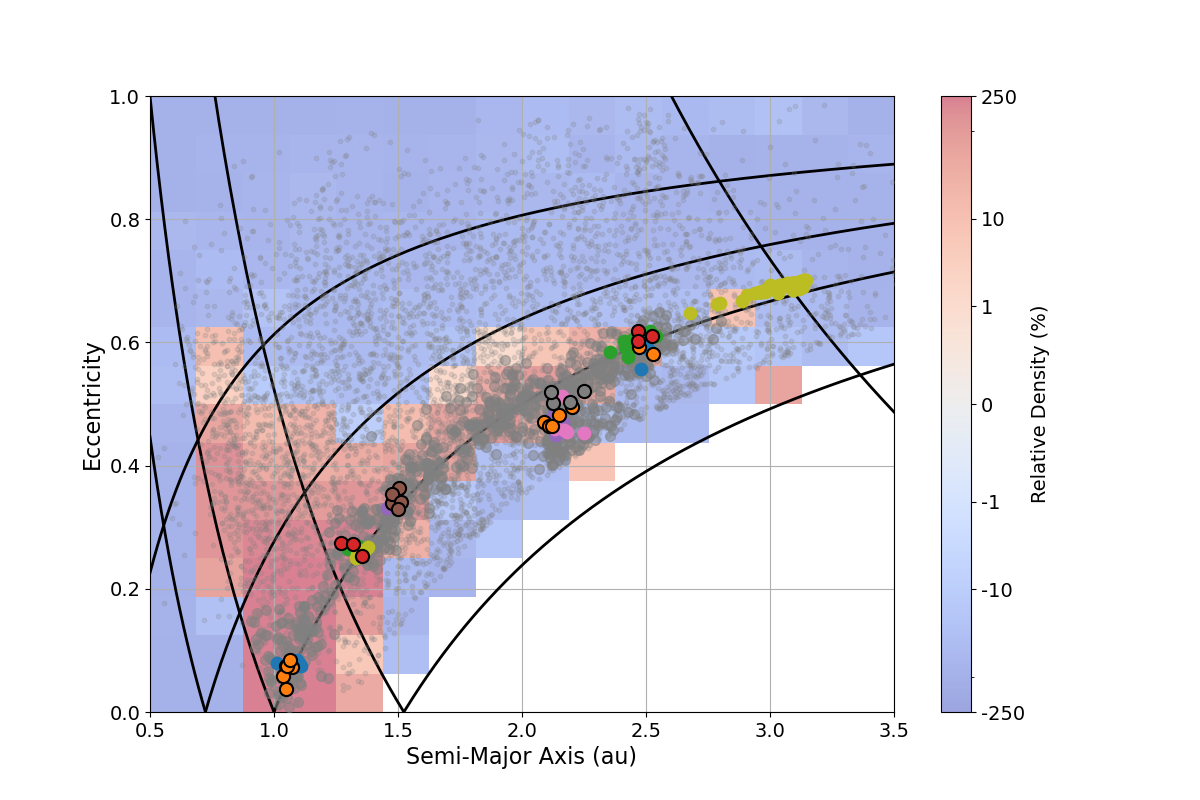}
        \label{fig:ae_cluster}
    \end{subfigure}%
    % \hspace{0.01\textwidth}
    \begin{subfigure}[b]{0.49\textwidth}
        \includegraphics[width=\textwidth]{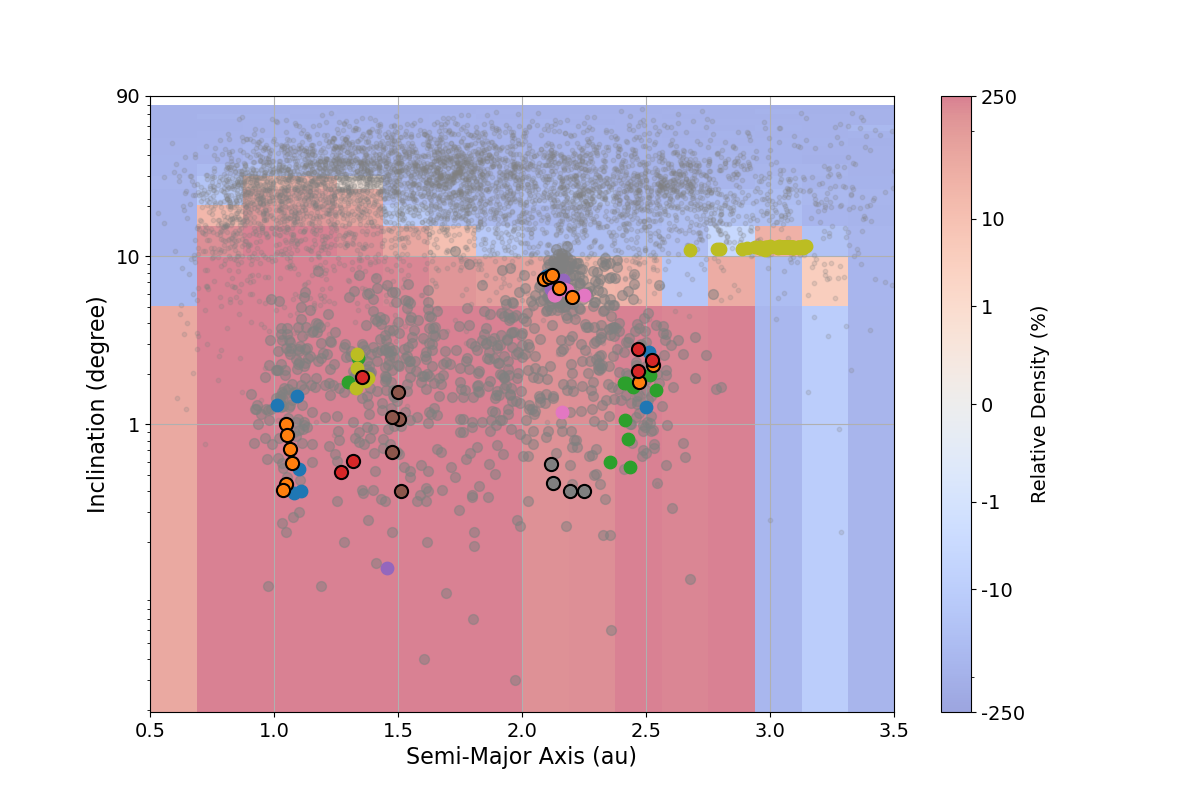}
        \label{fig:ai_clusters}
    \end{subfigure}%
    \caption{NEO orbital distribution with ephemeris taken from NASA Horizons with clusters identified, objects not belonging to a cluster can be seen as small gray points. The background heatmap, however, indicates the observational bias of the NEO dataset. The red regions denote orbits for which more objects exist in the NEO database relative to the debiased dataset of \citet{granvik2018debiased}, and blue is where there exists a deficit of NEO detections. Clusters identified by \citet{jopek2020orbital} can be seen as large dark gray points. The clusters identified within this study are colored, each with at least 5 members. They were identified using a DBSCAN algorithm where core points were defined as having at least two associations with a minimum $\epsilon$ corresponding to D$_{H}$ value of 0.03. This D-value was chosen as it corresponds well to where the cumulative D-value distribution displays a ``kink'' corresponding to an excess of similarity. The points that were also identified by \citet{jopek2020orbital} as being in a cluster have a black border. The new large yellow cluster identified is not asteroidal but corresponds to all the fragments of 73P/Schwassmann-Wachmann, which was a Jupiter-family comet observed to have undergone significant fragmentation between 1995-2006 \citep{reach2009distribution}.}
    \label{fig:NEO_clusters}
\end{figure*}

\begin{figure}[]
    \centering
    \begin{subfigure}[b]{0.5\textwidth}
        \includegraphics[width=\textwidth]{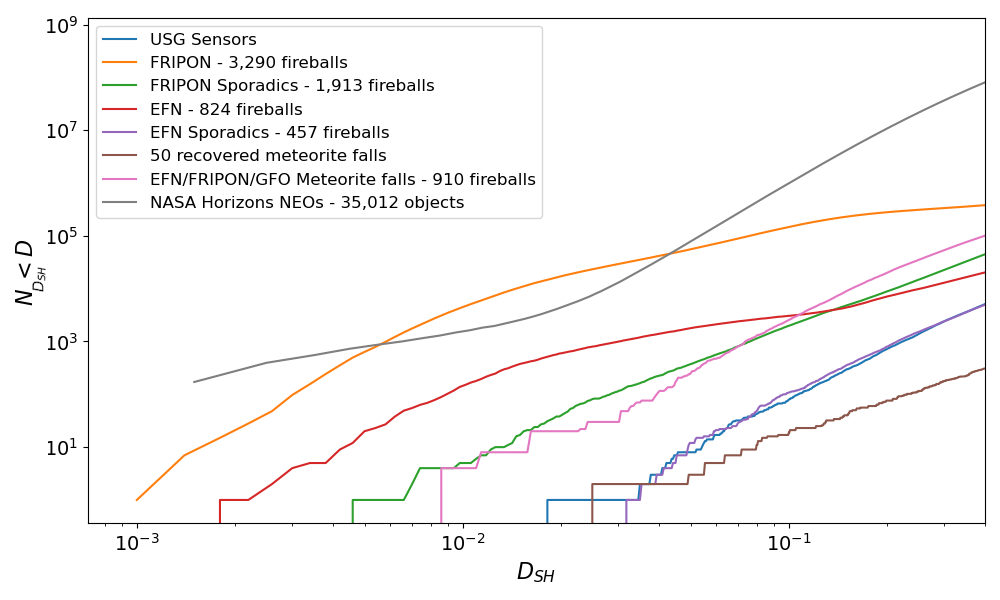}
        % \caption{$\iota$=0$^{\circ}$}
        \label{fig:combo_d_sh}
    \end{subfigure}
    
    \begin{subfigure}[b]{0.5\textwidth}
        \includegraphics[width=\textwidth]{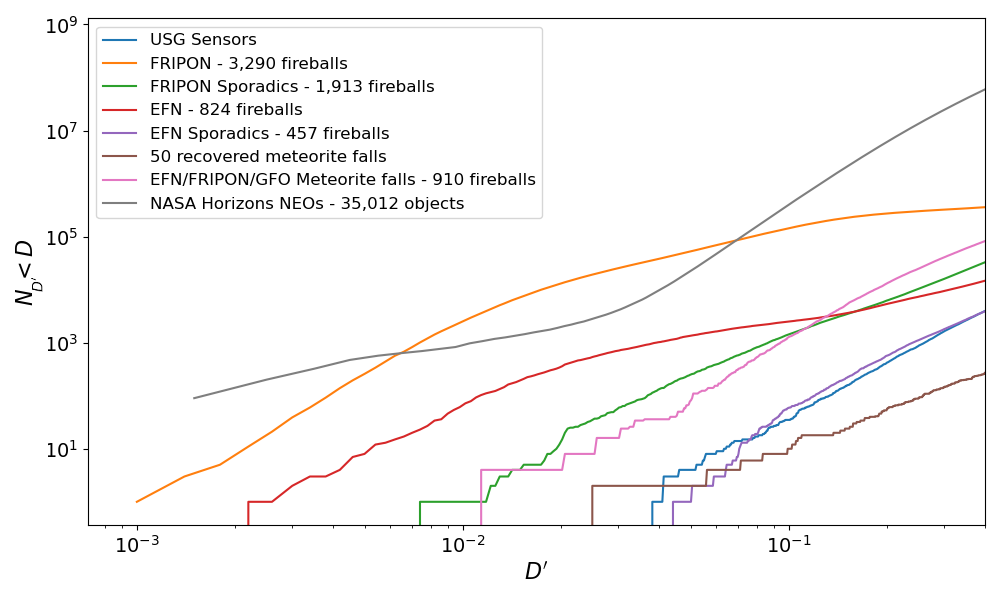}
        % \caption{$\iota$=10$^{\circ}$}
        \label{fig:combo_d_prime}
    \end{subfigure}

    \begin{subfigure}[b]{0.5\textwidth}
        \includegraphics[width=\textwidth]{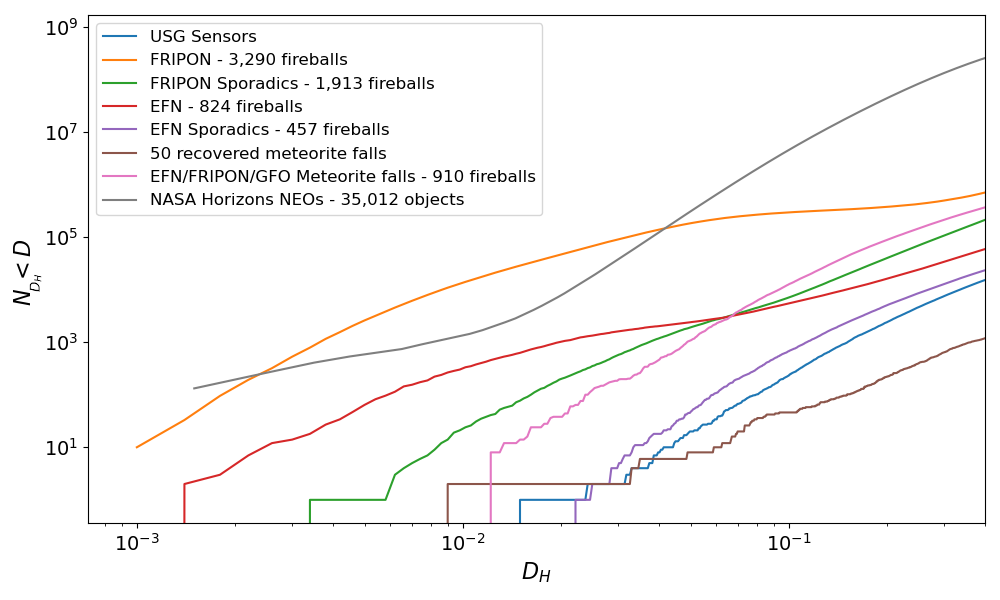}
        % \caption{$\iota$=20$^{\circ}$}
        \label{fig:combo_d_h}
    \end{subfigure}
    \caption{The cumulative D-value distributions for every possible pair within 310 USG Sensors orbits 3,290 FRIPON fireballs, 824 EFN fireballs, 1,913 FRIPON sporadic subset fireballs, 457 EFN sporadic subset fireballs, 50 recovered meteorite orbits, 616 potential 1\,g meteorite droppers observed by GFO/EFN/FRIPON, and 35,012 NEOs taken from the NASA Horizons ephemeris service.}
    \label{fig:all_combinations_slopes}
\end{figure}

We used the DBSCAN clustering algorithm employing D$_H$, an epsilon value of 0.03, and a minimum of two connections for a pair to be defined as a core point. This revealed clusters that tend to align near Earth-crossing orbits, as shown in Fig.~\ref{fig:NEO_clusters}. The epsilon value was chosen precisely where we see a ``kink'' in the cumulative D-value distribution (Fig.~\ref{fig:all_combinations_slopes}), indicating that a different production mechanism becomes dominant. This is just slightly larger than where we find the statistical significance is beyond 3-$\sigma$ for D$_{H}$ ($\sim$0.01), but significantly less than the minimum D$_H$ values used in \citet{jopek2020orbital}. \citet{jopek2020orbital} also uses a single-linkage clustering approach, which tends to produce elongated or "chain-like" clusters, especially in the presence of noise. The difference in clustering algorithm and larger D-value limits are likely the primary reasons why \citet{jopek2020orbital} identified hundreds of clustered NEOs compared to the tens that we identify here. That being said, we have not estimated the probability of having a group of pairings like done in \citet{jopek2020orbital}; thus, we are only plotting a minimum number of clusters where each linkage is statistically significant. 

In Fig.~\ref{fig:NEO_clusters}, we identified 12 clusters of at least 5 members satisfying our DBSCAN identification scheme. We found 66 clustered NEAs distributed between the 11 clusters, 28 of these objects being previously identified by \citet{jopek2020orbital}. These new identifications are mostly due to the significant increases in the NEO dataset size over the previous two decades. The NEO dataset used by \citep{jopek2020orbital} consisted of 20,032 NEAs, \citet{schunova2012searching} analyzed 7,563 NEOs, and the study by \citet{fu2005identifying} only had 3,319 NEOs to compare. Thus, the significantly larger dataset of 35,012 NEOs considered here is more amenable to discovering small ephemeral NEO families. However, this increase in NEO discoveries does come with its caveats, as surveys do not observe all objects equally. 

In Fig.~\ref{fig:NEO_clusters}, the heatmap shows the observational bias by comparing the synthetic debiased NEO dataset of \citet{granvik2018debiased} with the observed NEO dataset (35,012 objects). The red regions represent an overabundance of NEOs due to a discovery bias, while blue areas indicate under-representation in observations. Crucially, while we have demonstrated through our KDE Monte Carlo method that there is a statistically significant number of pairs on similar orbits as shown in Fig.~\ref{fig:false_positive_NEOS_NEOs}, all the clusters identified are in regions where an overabundance of NEOs is observed due to this discovery bias, as highlighted in Fig.~\ref{fig:NEO_clusters}. This does not necessarily imply that these clusters are entirely a result of the discovery bias, as our KDE Monte Carlo method should account for such biases, ensuring that the observed clustering is indeed statistically significant. However, the number of clustered objects identified by \citet{jopek2020orbital} is could be an overestimation based on our results, and this is possibly influenced by this observational bias. Notably, as seen in Fig.~\ref{fig:NEO_clusters}, there are regions, particularly on orbits with semi-major axes between 0.5 and 1.5\,au and higher eccentricities, where one would expect to see more clusters if they were solely due to observational bias. The absence of clusters in these regions supports our finding that while observational bias likely slightly contributes to the clustering identified, it cannot account for all of it. The clustering that we have identified also successfully includes the fragments of the well-observed breakup of comet 73P/Schwassmann-Wachmann (yellow group; Fig.~\ref{fig:NEO_clusters}) as a statistically significant cluster, distinct from the background, without prior intent for identification, giving us confidence that these clusters were identified using adequate constraints. 

This clustering observation is also consistent with recent studies suggesting that tidal disruptions are responsible for creating such groupings \citep{granvik2024tidal}. Our analysis identified approximately 10$^{1-2}$ such objects according to Fig.~\ref{fig:all_combinations_slopes}, supporting the hypothesis that tidal disruptions contribute to the observed clustering. These results interestingly contrast with orbital meteorites, fall data, and the CNEOS impact data set, for which no statistically significant pairings exist. This disparity between the NEO groupings and the meteorite and impact data remains an intriguing point for further investigation, particularly in understanding the differing influences of tidal disruptions and other dynamic processes on these populations. Other mechanisms, such as rotational disruptions, meteoroid impacts, etc., are other possible debris generation mechanisms worthy of exploring, particularly for smaller debris \citep{jewitt2012active}.  

%%%%%% talk about trying to cluster all 616 1g falls and 50 meteorites with the 35,012 NEOs.... Absolutely none found even for just clusters of 3 with NEOs ... why? bias, less in fireball database... like 0.1% ? ?? Could it be due to just the intrinsic higher uncertainties (somewhat even more than acknowledged...) of fireball data.... could this effectively just wipe out any possibility of detecting such small streams? I think so... I think that is it... the uncertainties of fireballs are acting like a natural KDE 

Given the clustering observed in the NEO dataset, which accounts for up to a few percent of the population \citep{jopek2020orbital}, a similar proportion of clustering would be expected in the fireball and impact datasets if these populations are comparable. To explore this hypothesis, we also integrated the fireball data with the NEO dataset and reapplied the DBSCAN clustering algorithm utilized in generating our NEO clusters Fig.~\ref{fig:NEO_clusters}. Surprisingly, this combined analysis revealed no changes in the identified clusters; the same clusters persisted in the NEO dataset without any inclusion of fireball data. This unexpected result suggests two possibilities: either the inherent uncertainties in fireball data are effectively smoothing out potential small associations, or the proportion of clusters within the impact dataset is actually lower than what we observe within the NEO population.

This decrease in the proportion of detectable clusters could also be influenced by the intrinsic biases of the two datasets. The NEO dataset and the impact dataset are both biased, but in opposite ways. The NEO dataset is skewed towards detecting larger and brighter objects on orbits near the Earth (Fig.~\ref{fig:NEO_clusters}), which tend to have lower encounter velocities and more Earth-like orbits. On the other hand, the fireball dataset is biased towards higher velocity impacts, particularly those from smaller objects, because higher velocity impacts tend to be brighter and thus more likely to be observed. Consequently, this dataset is more likely to capture events with larger semi-major axes, higher eccentricities, and greater inclinations, the opposite of the bias seen in the NEO dataset. Therefore, these conflicting biases could also contribute to the lack of observed associations between the fireball data and the NEO clusters.

To extend this, we have also compared the observed cumulative D-distributions (i.e., the distribution of all distance values between all objects in the population(s)). This was visualized in a log-log plot in Fig.~\ref{fig:all_combinations_slopes}, where the y-axis represents the number of objects with a D-value less than that on the x-axis. We observed that datasets showing no evidence of statistically significant associations exhibited a linear distribution in log-log space for this cumulative plot. We calculated the orbital similarity of all unique pair combinations within the USG sensor data, the 50 recovered meteorite falls, 616 fireballs observed by GFO/FRIPON/EFN, and the sporadic subsets of the FRIPON and EFN datasets after removing known meteor showers using the Valsecchi D-function \citep{valsecchi1999meteoroid}. This function was only used to remove the shower component as it does not rely on orbital similarity directly, avoiding the self-fulfilling bias of the cumulative d-value slope analysis. The linearity in log-log space suggests a lack of significant streams, as there is no excess of pairs with small D-values to indicate clustering beyond what is expected by random chance. This linear behavior can be explained by the random distribution of orbital elements. In contrast, datasets with significant streams show a deviation from this linear trend, with a change in slope at smaller D-values, indicating an excess of pairs with high orbital similarity. This is demonstrated clearly through the contrast between the FRIPON and EFN complete datasets and the sporadic sub-component.

Our analysis indicates that the populations without evidence of streams, including the USG sensors and the 50 recovered meteorite falls, follow this linear trend for a cumulative D-value distribution. This provides further evidence that no significant meteoroid streams are present in these datasets. The lack of deviation from linearity in the log-log cumulative plots confirms that the observed orbital similarities are consistent with random associations, reinforcing the conclusions drawn from our KDE-based Monte Carlo simulations. The cumulative D-value distributions for the sporadic populations exhibit very similar slopes, suggesting a general degree of similarity between their distributions and, thus, self-similarity. Notably, the 50 meteorite falls deviate from this pattern by displaying a slightly lower slope while maintaining linearity. This lower slope does not resemble the FRIPON and EFN datasets with showers included. Additionally, it contrasts with the observed fall database of 616 potential meteorite-dropping fireballs observed by the GFO, EFN, and FRIPON. This difference likely indicates a discovery or recovery bias influencing our recovered meteorite database. This is not entirely surprising, given the challenges and limitations of meteorite recovery. 
%This bias has been confirmed and quantified in the recent analysis of \citet{}. 

Notably, the NEO dataset, as previously discussed, does display a significant slope-change or ``kink'' in the cumulative d-value distribution around 10$^{-2}$. This clear change in slope supports the hypothesis that tidal disruption or another mechanism in near-Earth space generates short-lived small NEO families \citep{schunova2014properties,granvik2024tidal}. However, the absence of a significant change in slope in the cumulative D-value plots across all other analyzed datasets corroborates our findings that there is no substantial evidence for meteoroid streams among meteorite falls or CNEOS impacts. 

% ADD A CLUSTER PLOT

% add in cumulative sum log-log plot of D values, showing ones with significant clustering and streams 
% contributions have a bend in the slope 

\subsection{Lyapunov Characteristic Lifetimes}
% Need to add a transitionary paragraph here... 
Building upon the analysis of orbital similarities and the identification of statistically significant clusters within the NEO population, we now turn our attention to the underlying dynamical behavior of these objects. The chaotic nature of near-Earth space, influenced heavily by gravitational interactions, suggests that the orbital paths of these objects are not only influenced by their current configuration but are also highly sensitive to initial conditions. To explore this, we investigated the Lyapunov characteristic lifetimes of NEO orbits, which quantitatively measure their dynamic stability. Specifically, we examine how these lifetimes vary with orbital parameters such as semi-major axis, eccentricity, and inclination, focusing on their implications for the temporal evolution of meteoroid streams and the practical limits of using orbital similarity metrics like the D-values over extended periods.

\begin{figure}[!htbp]
    \centering
    \begin{subfigure}[b]{0.5\textwidth}
        \includegraphics[width=\textwidth]{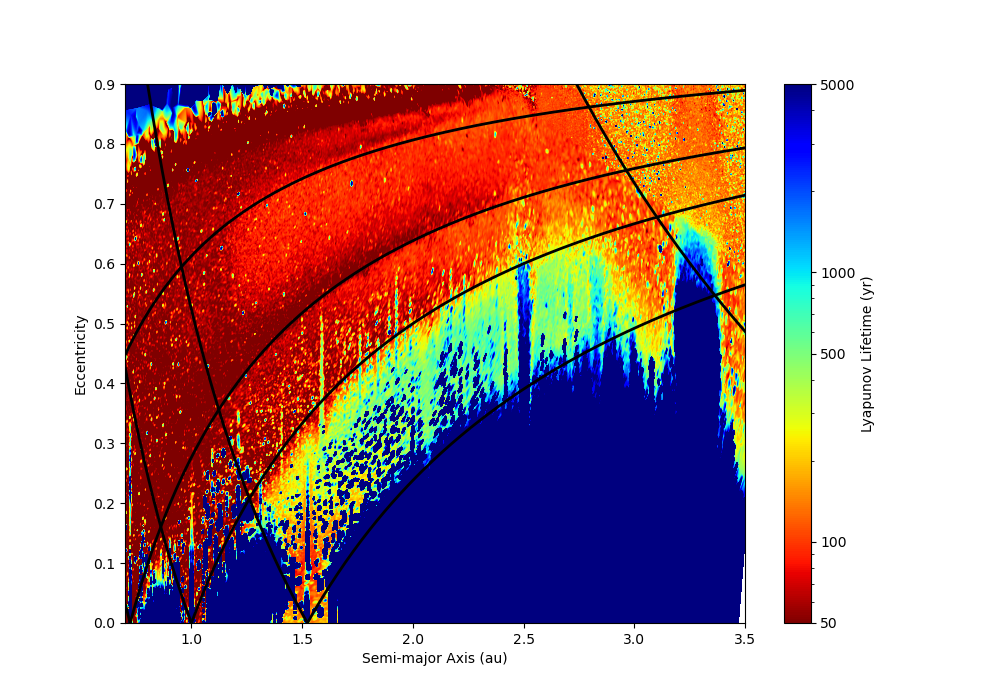}
        \caption{$\iota$=0$^{\circ}$}
        \label{fig:moid_lyapunov_JFCs}
    \end{subfigure}
    
    \begin{subfigure}[b]{0.5\textwidth}
        \includegraphics[width=\textwidth]{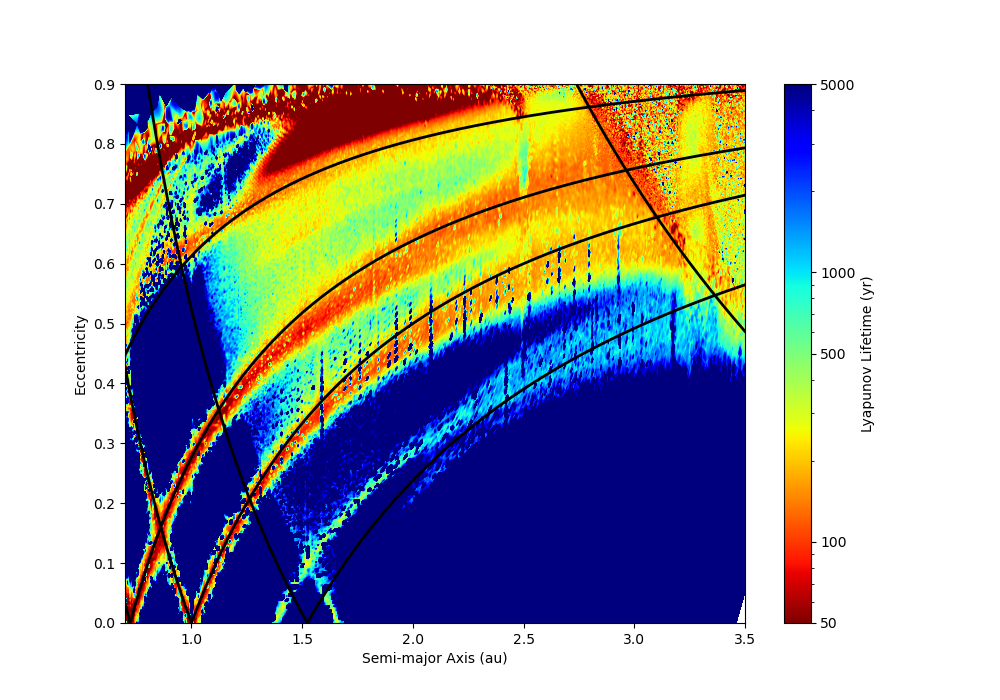}
        \caption{$\iota$=10$^{\circ}$}
        \label{fig:moid_lyapunov_DFN}
    \end{subfigure}

        \begin{subfigure}[b]{0.5\textwidth}
        \includegraphics[width=\textwidth]{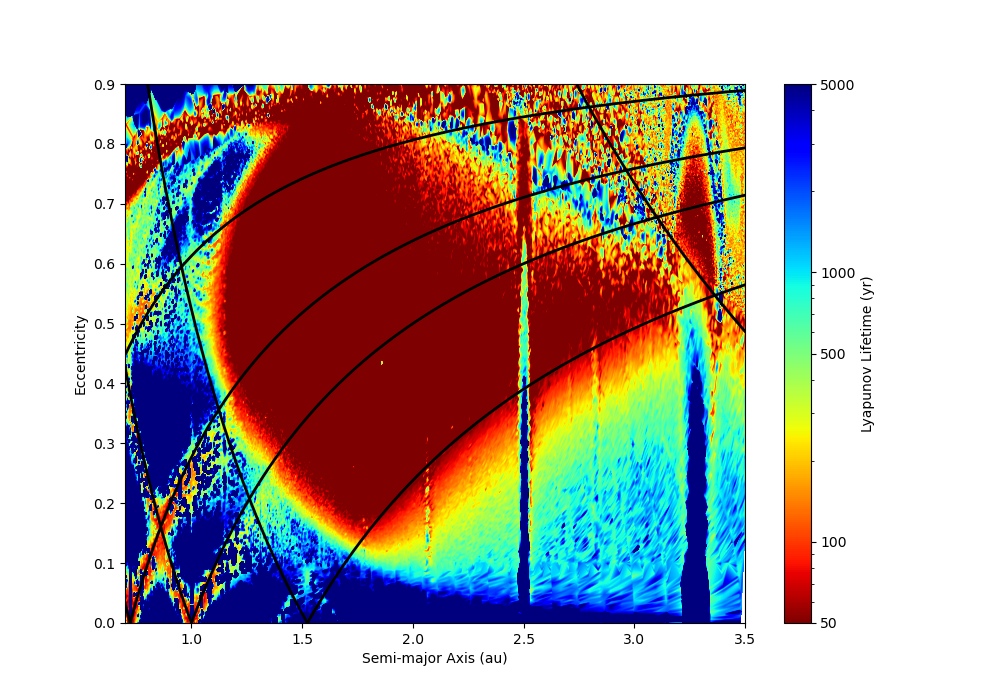}
        \caption{$\iota$=20$^{\circ}$}
        \label{fig:moid_lyapunov_EFN}
    \end{subfigure}

    \caption{Lyapunov characteristic lifetime as a function of the semi-major axis and eccentricity with an inclination of 0$^{\circ}$ (a), 10$^{\circ}$ (b), 20$^{\circ}$ (c). The black lines indicate the where the aphelion/perihelion distances are equal to the nominal semi-major axis value of Mercury, Venus, Earth, Mars, or Jupiter.}
    \label{fig:lyapunov_maps}
\end{figure}

Figure~\ref{fig:lyapunov_maps} presents the Lyapunov characteristic lifetimes as functions of the semi-major axis and eccentricity for three distinct inclinations: 0$^{\circ}$, 10$^{\circ}$, and 20$^{\circ}$. These maps were produced using the Rebound N-body simulation software. The model of the solar system included the Sun and the eight major planets, utilizing initial conditions sourced from JPL Horizons. The 'whfast' integrator, known for its efficiency with near-Keplerian orbits, was employed with a time step set at 0.01 years \citep{rein2015whfast}.

A dense grid of points represents initial orbital parameters for hypothetical small bodies for each map with linear interpolation in lower statistic regions. Specifically, 289,640 particles were integrated at 0$^{\circ}$ inclination, 287,352 at 10$^{\circ}$ inclination, and 353,470 particles at 20$^{\circ}$ inclination, covering a semi-major axis range from 0.5 to 3.5\,au and eccentricities from 0 to 0.95. The color scale in each map indicates the Lyapunov lifetime, which quantifies the rate at which two infinitesimally close orbits diverge. This highlights the system's sensitivity to initial conditions. The simulations were run over a period of 20,000 years, and the Lyapunov time was calculated as the inverse of the Lyapunov exponent.

The Lyapunov lifetime maps exhibit significant variation across different orbital elements and inclinations. At a low inclination of 0$^{\circ}$, the maps display considerable variability in lifetimes, with shorter times noted in regions of higher eccentricity as they cross the orbits of more planets. Notable, too, are mean-motion resonances such as the 3:1 and 2:1 with Jupiter, which appear as zones of enhanced stability on these timescales (longer Lyapunov lifetimes). This is linked to close encounters:  meteoroids inside them are unable to encounter the planet they are in resonance with. With an increase to a moderate inclination of 10$^{\circ}$, the chaotic zones tend to be further concentrated towards the aphelion/perihelion lines of the planets as these bodies tend to have close encounters. At the highest studied inclination of 20$^{\circ}$, chaotic zones are more widespread, especially for orbits intersecting multiple planetary paths. The protective effect of mean-motion resonances lessens, and overall stability across the semi-major axis range decreases, suggesting that higher inclinations generally lead to more chaotic orbits in near-Earth space (except for more circular orbits). This protection of the mean-motion resonance is particularly not visible for large eccentricities when the inclination reaches 20$^{\circ}$. A previous work \citep{Courtot_al_2024} has shown that a chaos indicator in that region reveals regular dynamics. This could be explained in two ways: this area could be qualified as "stable chaos," or the chaotic aspect could also appear only on longer timescales, such as the 20,000 years used here.

% The maps in Fig.~\ref{fig:lyapunov_maps} show that the more planets-crossing, the shorter the Lyapunov lifetime. It seems only Jupiter is excepted from this rule, since areas corresponding to encounters with Jupiter (on top of encounters with other planets) show a longer Lyapunov lifetime than expected. This is surprising as Jupiter is often responsible for disruptions in the streams \citep{Egal_al_2019}. This could be an indication of stable chaos, where the Lyapunov lifetimes would point to regular dynamics, whereas a chaos indicator would not. 

Comparing the three maps with each other, the evolution of the Lyapunov lifetime as the inclination rises seems to be strongly linked with close encounters. As the meteoroids are further away from the ecliptic, the zones of shortest Lyapunov time change shape and follow less clearly the zones of close encounters delimited by the black lines. It is well-known that the geometry of close encounters, and thus its effects on meteoroid streams, depends on the direction of the incoming meteoroid with respect to the planet's orbit \citep{Carusi_al_1990}. This explains how the shape of areas of shorter Lyapunov lifetime can be distorted for higher inclinations. We can also note that for the inclination at 10$^{\circ}$, therefore close to Mercury's orbital inclination at 7.004$^{\circ}$, Lyapunov lifetimes are especially short when close encounters with Mercury are possible, meaning Mercury would be the driving force behind the chaos in this configuration. 

For the inclination at 0$^{\circ}$, a large portion of the map includes Lyapunov lifetimes shorter than 100 years, while this value can go down to 60 years for a large part of the map for the inclination at 20$^{\circ}$. The map at 10$^{\circ}$ shows less areas with short Lyapunov lifetime, except for encounters with Mercury, as discussed above. Therefore in those regions, meteoroids with initially similar orbits might diverge quickly, rendering the use of D-functions quite complex. A short Lyapunov lifetime reveals areas where meteoroids on initially similar orbits might reach widely different orbits. Even if D-values are able to characterize whether the orbits are similar today, it does not predict how they will evolve: an existing stream today might disappear quite fast. Therefore, a short Lyapunov lifetime in an area only shows where any D-function results must be taken with a grain of salt.

Finally, it should be noted that we are especially interested in the area where meteoroids can encounter the Earth, since D-fucntions are mostly used to find meteor showers. Taking this condition into account, the Lyapunov lifetime is usually very short (less than 200~years for $i = 0^{\circ}$, less than 500~years for $i = 10^{\circ}$ and less than 100~years for $i = 20^{\circ}$), and it gets much longer on only very small areas. This means that for the conditions we are especially interested in, results from D-functions should be regarded with caution.

The findings from the Lyapunov lifetime maps have significant implications for the interpretation of statistical clusters in near-Earth space, as discussed in the previous section. Despite the presence of tidal disruptions in near-Earth space, the orbital paths that result from these disruptions also tend to be the most chaotic regions. The clusters identified in Fig.~\ref{fig:NEO_clusters}, correspond to Lyapunov characteristic lifetimes of only tens to hundreds of years — as the perihelia values tend towards 1\,au. Any clustering of objects resulting from recent tidal disruption events would have a very short dynamical memory. Consequently, while we can statistically analyze orbital similarities among near-Earth objects, precisely identifying which objects belong to a particular cluster becomes problematic due to the rapid divergence of orbits within these chaotic regions.

This challenge is further underscored by the contrast between the frequency of tidal disruption events and the typical Lyapunov lifetimes. For instance, the estimated frequency of tidal disruptions with the Earth that would result in a detectable cluster in the NEO population has been estimated previously to be anywhere between a few to several thousand years \citep{richardson1998tidal,toth2011tidal,schunova2014properties}. Thus, while tidal disruption could generate clusters, the chaotic nature of their orbits means that meaningful connections between the bodies in these clusters would disperse very quickly. In conjunction with higher intrinsic uncertainties of orbits extracted from short observation arcs as a body impacts the atmosphere, this chaotic dispersion may also help explain why we do not observe statistically significant clustering in fireball datasets. With increased fireball data, clusters may become statistically significant, but perhaps this will require thousands to tens of thousands of very precise fall observations. Currently, the lack of statistically significant clustering in fireball data suggests that any observed similarities are more likely to be spurious associations rather than evidence of real clustering. Any current claims linking specific meteorites to near-Earth objects should be viewed skeptically. The statistical analysis presented here indicates that such associations are not statistically sound and are more likely to result from random chance rather than genuine physical connections.

\subsection{Meteoroid Stream Decoherence Lifetimes}
% talk about how decoherence lifetime differs from the chaos indicator 
While the Lyapunov characteristic lifetime offers valuable insights into the chaotic nature of individual meteoroid orbits, it does not fully encapsulate the persistence of coherent structures within a meteoroid stream. The Lyapunov lifetime typically reflects the timescale over which small perturbations lead to exponential divergence in the orbits of individual particles, usually on the order of a few hundred years in near-Earth space. This measure primarily addresses the evolution of one particle compared to others with similar initial conditions rather than the longevity of the collective behavior of a meteoroid stream. In contrast, the decoherence lifetime of a meteoroid stream focuses on the collective dynamics of the stream's constituent particles, quantifying the duration over which the stream remains identifiable as a distinct, statistically significant entity, despite the underlying chaotic dynamics. This lifetime often spans tens of thousands of years, significantly longer than the Lyapunov lifetime, because it considers the overall correlation in the orbits of the group of meteoroids, which can remain similar enough to maintain the stream's identity over extended periods.

A potentially helpful analogy to illustrate this difference is to think of a cloud of smoke rising from a chimney. As the smoke particles ascend, small air currents cause them to move in different directions, leading to the rapid dispersion of individual particles. The point at which the paths of individual smoke particles start diverging from each other is analogous to the Lyapunov lifetime, where minor differences in initial conditions lead to rapid divergence in the orbits of meteoroids. However, despite the individual particles dispersing, the cloud itself remains visible as a cohesive entity for a considerable time as it drifts away from the chimney. The overall shape and presence of the smoke cloud linger, even as individual particles follow chaotic paths. This extended visibility and coherence of the cloud represent the decoherence lifetime, during which the smoke continues to be recognized as a collective stream despite the chaos experienced by individual particles. This analogy helps to clarify how a stream can remain a discernible group long after its constituents have started to exponentially diverge.

% talk about how we chose our decoherence definition (how did Pauls & Gladman do it?) 
The study by \citet{pauls2005decoherence} estimated the decoherence lifetimes for the three hypothetical meteoroid streams—Příbram, Innisfree, and Peekskill — also by generating test particles with nearly identical Keplerian orbits. These particles were integrated 500\,kyrs considering the gravitational influences of all nine planets. However, since they only considered three streams, they did not need to place a strict condition on when decoherence officially occurred. They defined the stream as decoherent when the orbits of the particles had spread out sufficiently that any tight clustering in their orbits, which would have indicated a coherent stream, had disappeared. This loss of coherence was monitored by observing the dispersion of the orbits over time, specifically by examining how the encounter orbital longitudes of particles with Earth spread out. 

In our study, we sought to define the decoherence lifetimes of meteoroid streams in a manner that would allow us to systematically compare the persistence of these streams across different regions of orbital space. This was important as a more qualitative definition of decoherence was impractical, given that we had generated 300 fictitious streams in total to achieve this goal. To do this, we considered the stream to have become ``decoherent'' when the largest remaining cluster within the stream only constituted 5\% or less of the original size of the stream. This threshold was selected to ensure that we were capturing the significant dispersion of the stream beyond mere perturbations, reflecting a point where the stream's identity has effectively been lost. We once again used the DBSCAN clustering algorithm, as it allowed us to identify and track clusters over time. 

The choice of $\epsilon$ values in our DBSCAN clustering analysis is critically important, as it directly influences how we define and detect the coherence of meteoroid streams. The $\epsilon$ value determines the maximum distance between two points to be considered part of the same cluster, effectively setting the sensitivity threshold for identifying coherent structures within the data. The key challenge here is that the appropriate $\epsilon$ value is not constant; it must be carefully adjusted based on the size of the dataset and the proportion of the dataset that is made up of meteoroid streams. When streams constitute a small fraction of the overall data, smaller $\epsilon$ values are required to avoid mistakenly merging distinct groups or failing to detect smaller streams. Conversely, larger $\epsilon$ values can be used in datasets where streams form a more significant portion to capture the broader structures and avoid fragmenting real streams into smaller, artificially distinct clusters.

This dependency is illustrated in Fig.~\ref{fig:all_combinations_slopes}, where we observe distinct changes in the slope of the cumulative D-value distributions for the FRIPON and EFN databases compared to the NEO dataset. The D-values where these changes occur are approximately an order of magnitude larger for the FRIPON and EFN databases than for the NEO dataset. This difference is primarily due to the higher proportion of cometary meteoroids in the FRIPON and EFN data, which make up a significant fraction of the observed orbits. Specifically, as shown in \cite{shober2024generalizable}, meteor showers constitute roughly 18-25\% of the EFN dataset. In contrast, streams in the NEO dataset represent only a small fraction, possibly just a few percentage points at most \citep{jopek2020orbital}. This disparity necessitates a larger $\epsilon$ for the FRIPON and EFN databases to appropriately cluster the significant stream components. In contrast, with its more dispersed and sparse stream population, the NEO dataset requires a much smaller $\epsilon$ to identify the subtle clustering signals.

Our specific choice of $\epsilon$ values was thus strategically tied to the underlying processes we aimed to study, given the specific populations at hand. The $\epsilon$ value for the DBSCAN algorithm was chosen based on the ``kink'' observed in the cumulative distribution of the D-values for NEOs (Fig.~\ref{fig:all_combinations_slopes}). We interpret this unique slope change as a signature of tidal disruptions -- a mechanism distinct from the cometary outgassing that dominates the FRIPON and EFN datasets. By aligning our $\epsilon$ selection with this kink, we targeted the detection of streams generated through tidal disruptions, allowing us to map the decoherence lifetimes of these streams within near-Earth space. This approach ensures that the resulting decoherence maps are representative of the decoherence lifetimes of streams generated from tidal descriptions or any other mechanism that is causing this excess of similarity amongst NEOs. We interpret this ``kink'' as being likely linked to tidal disruptions, as this phenomenon has been supported by recent studies \citep{schunova2014properties,granvik2024tidal}. These are the same $\epsilon$ values as those chosen in the cluster identifications of Fig.~\ref{fig:NEO_clusters}. 

% talk about decoherence as a function of space and criterion 

\begin{figure}[]
    \centering
    \begin{subfigure}[b]{0.5\textwidth}
        \includegraphics[width=\textwidth]{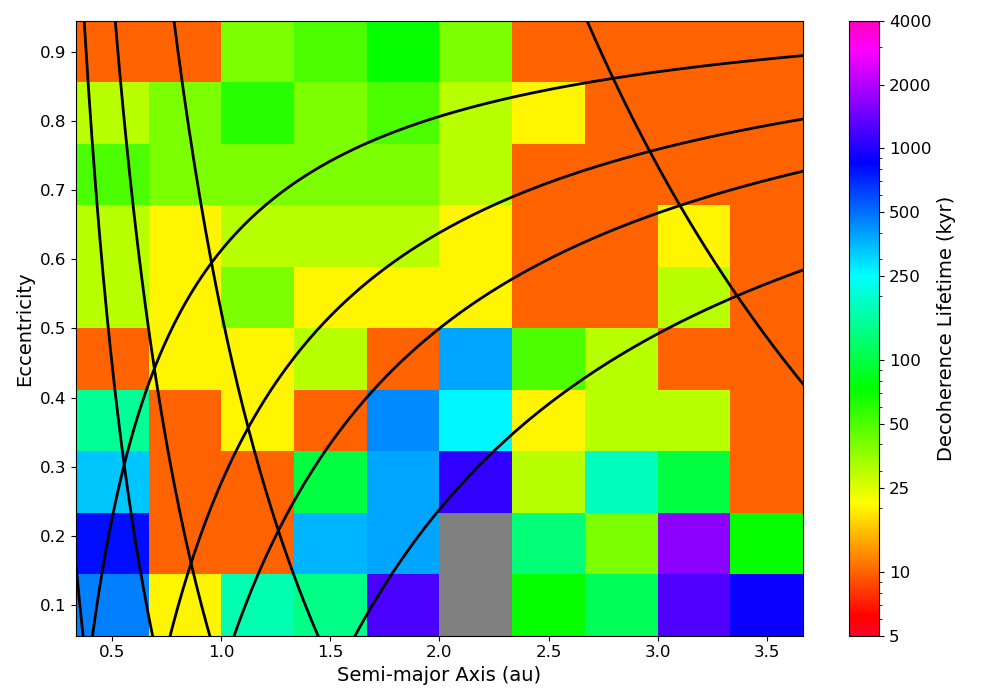}
        \caption{$D_{SH}$}
        \label{fig:d_sh_decoherence_0deg}
    \end{subfigure}
    
    \begin{subfigure}[b]{0.5\textwidth}
        \includegraphics[width=\textwidth]{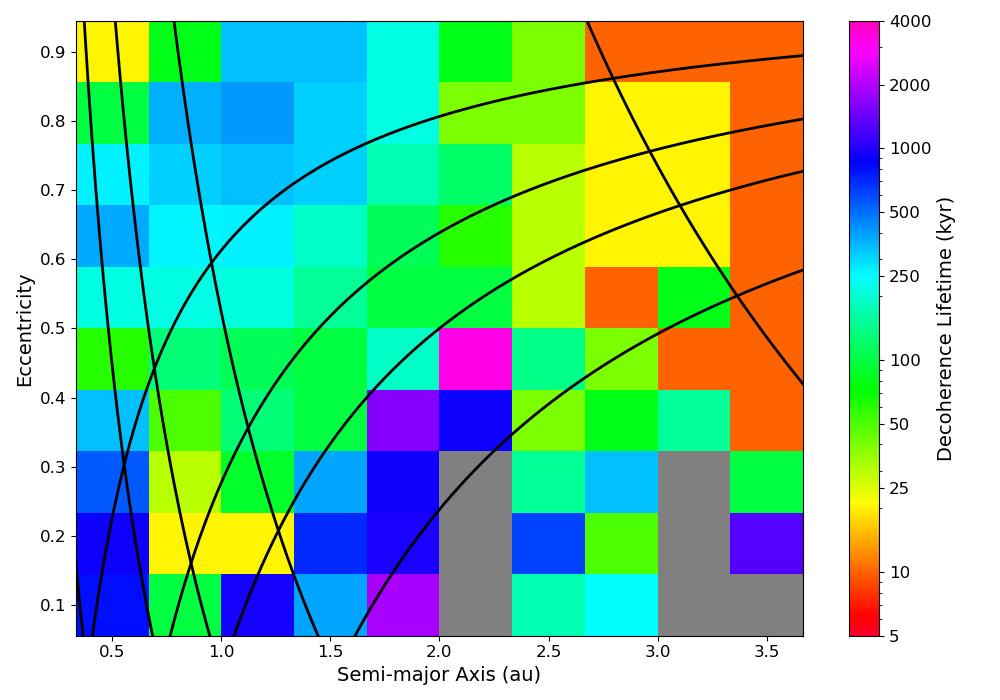}
        \caption{$D'$}
        \label{fig:d_prime_decoherence_0deg}
    \end{subfigure}

    \begin{subfigure}[b]{0.5\textwidth}
        \includegraphics[width=\textwidth]{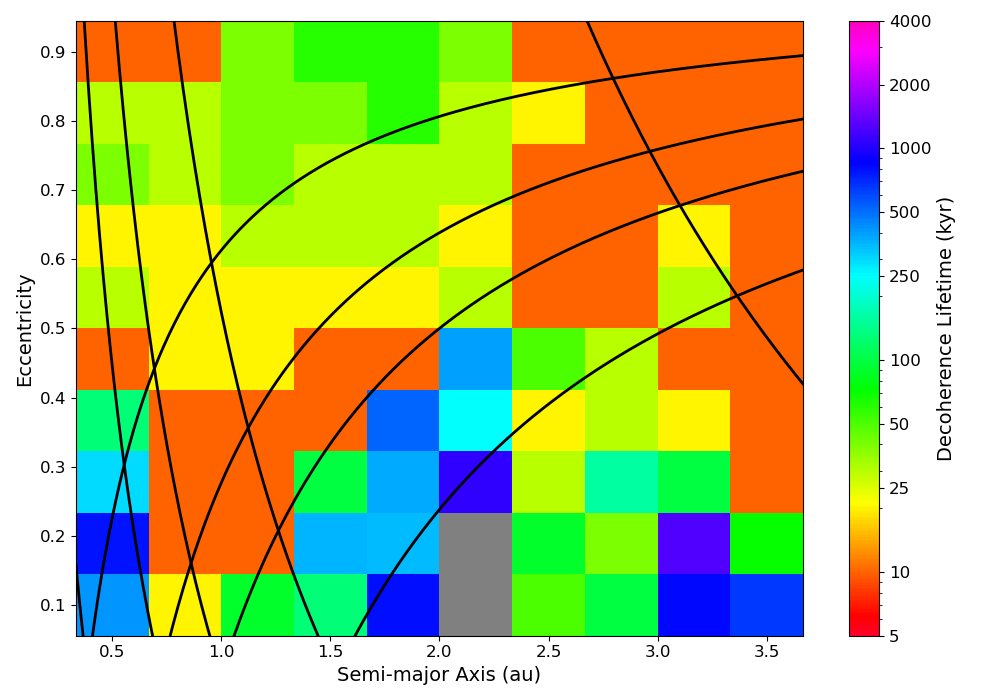}
        \caption{$D_{H}$}
        \label{fig:d_h_decoherence_0deg}
    \end{subfigure}

    \caption{Decoherence lifetime as a function of the semi-major axis and eccentricity with an inclination of 0$^{\circ}$ using $D_{SH}$ (a), $D'$ (b), or $D_{H}$ (c) values in conjunction with a DBSCAN algorithm to determine clustering. The $\epsilon$ value for the DBSCAN algorithm was chosen to where the ``kink'' is in the cumulative D-value distributions for NEOs (Fig.~\ref{fig:all_combinations_slopes}). Also, core points have at least 2 connections. The decoherence lifetime was thus defined as when the fictitious meteoroid stream lost 95\% or more of the original stream mass.}
    \label{fig:0deg_decoherence_maps}
\end{figure}

\begin{figure}[]
    \centering
    \begin{subfigure}[b]{0.5\textwidth}
        \includegraphics[width=\textwidth]{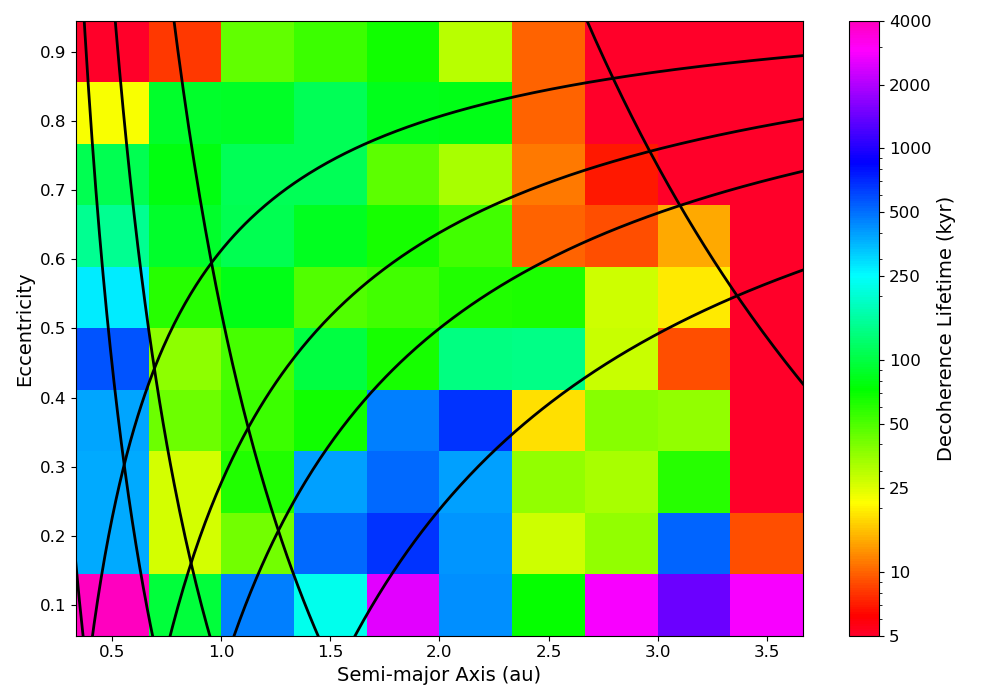}
        \caption{$D_{SH}$}
        \label{fig:d_sh_decoherence_10deg}
    \end{subfigure}
    
    \begin{subfigure}[b]{0.5\textwidth}
        \includegraphics[width=\textwidth]{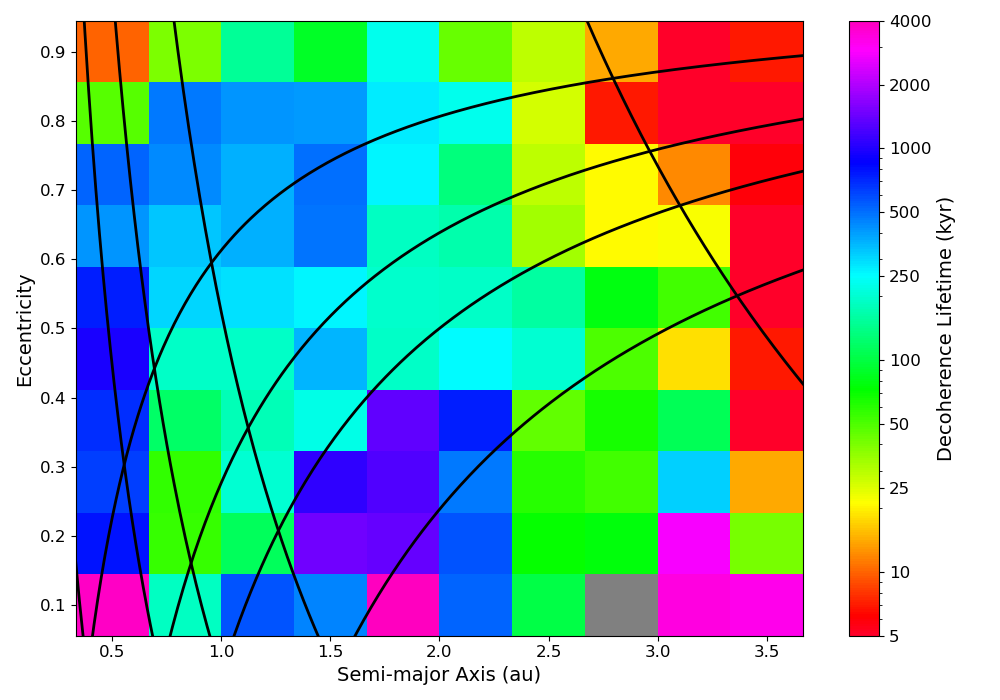}
        \caption{$D'$}
        \label{fig:d_prime_decoherence_10deg}
    \end{subfigure}

    \begin{subfigure}[b]{0.5\textwidth}
        \includegraphics[width=\textwidth]{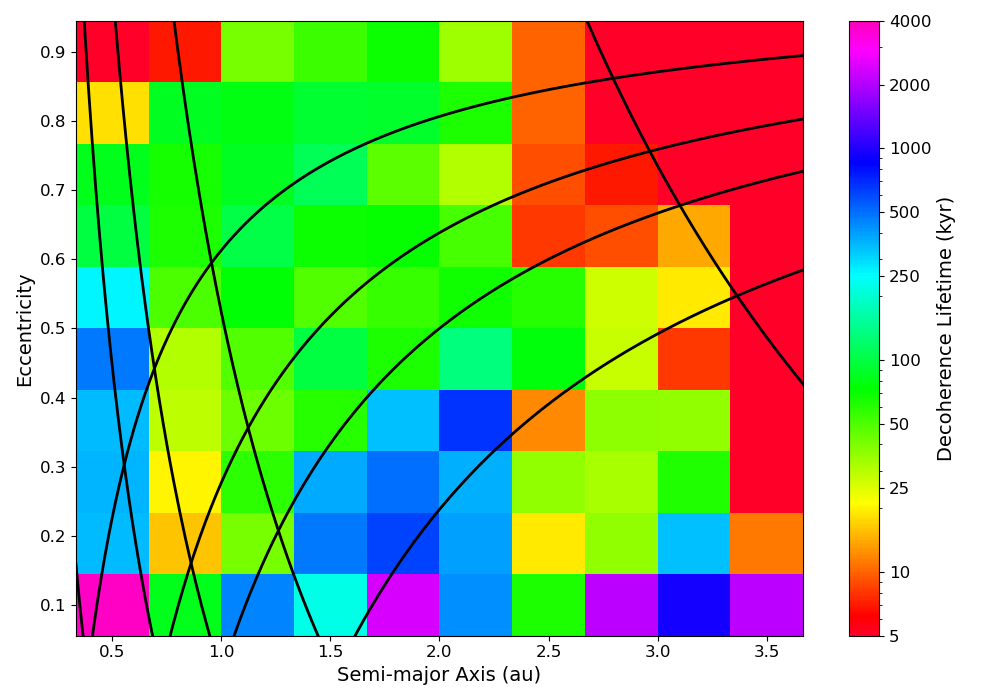}
        \caption{$D_{H}$}
        \label{fig:d_h_decoherence_10deg}
    \end{subfigure}

    \caption{Decoherence lifetime as a function of the semi-major axis and eccentricity with an inclination of 10$^{\circ}$ using $D_{SH}$ (a), $D'$ (b), or $D_{H}$ (c) values in conjunction with a DBSCAN algorithm to determine clustering. Decoherence time was defined as when a fictitious meteoroid stream lost 95\% or more of the original stream.}
    \label{fig:10deg_decoherence_maps}
\end{figure}

\begin{figure}[]
    \centering
    \begin{subfigure}[b]{0.5\textwidth}
        \includegraphics[width=\textwidth]{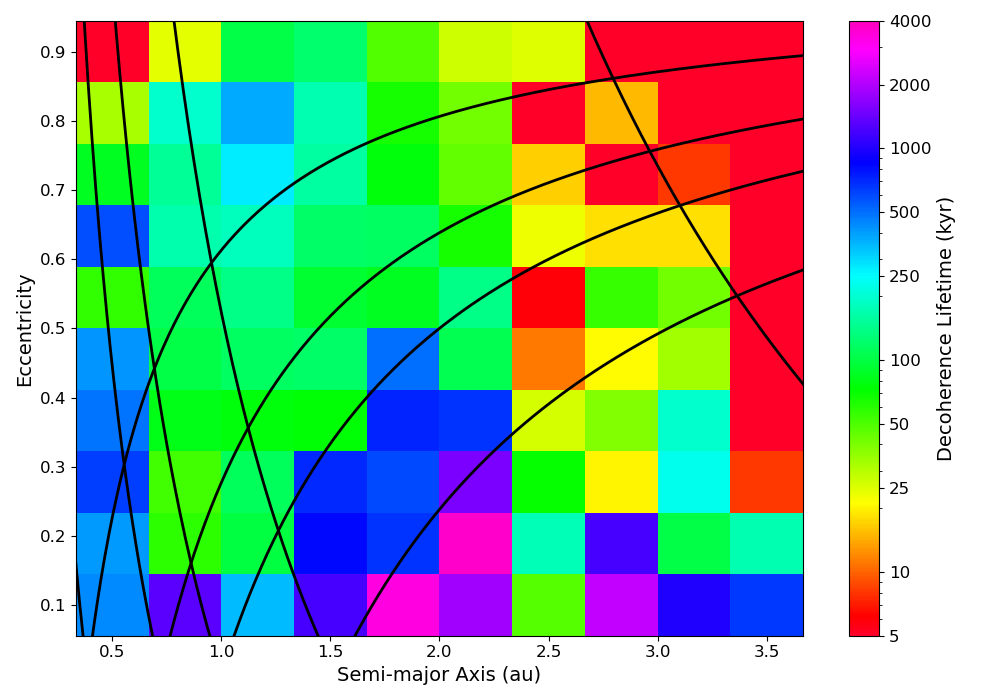}
        \caption{$D_{SH}$}
        \label{fig:d_sh_decoherence_20deg}
    \end{subfigure}
    
    \begin{subfigure}[b]{0.5\textwidth}
        \includegraphics[width=\textwidth]{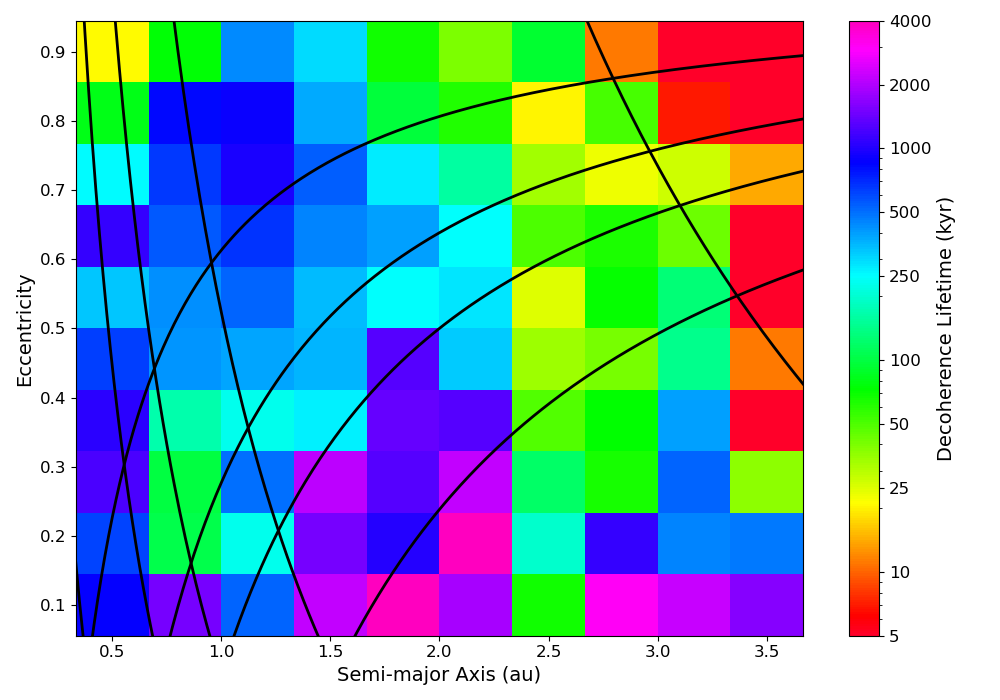}
        \caption{$D'$}
        \label{fig:d_prime_decoherence_20deg}
    \end{subfigure}

    \begin{subfigure}[b]{0.5\textwidth}
        \includegraphics[width=\textwidth]{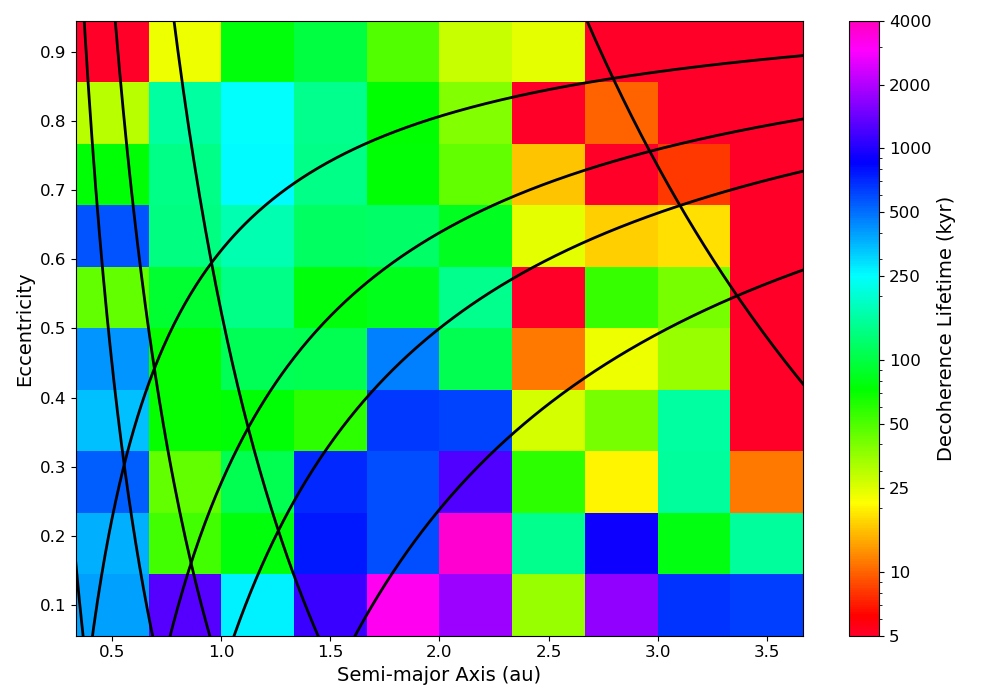}
        \caption{$D_{H}$}
        \label{fig:d_h_decoherence_20deg}
    \end{subfigure}

    \caption{Decoherence lifetime as a function of the semi-major axis and eccentricity with an inclination of 20$^{\circ}$ using $D_{SH}$ (a), $D'$ (b), or $D_{H}$ (c) values in conjunction with a DBSCAN algorithm to determine clustering. Decoherence time was defined as when a fictitious meteoroid stream lost 95\% or more of the original stream.}
    \label{fig:20deg_decoherence_maps}
\end{figure}

The resulting decoherence maps, presented in Fig. \ref{fig:0deg_decoherence_maps}, \ref{fig:10deg_decoherence_maps}, and \ref{fig:20deg_decoherence_maps}, illustrate how these lifetimes vary as a function of orbital space and the similarity criterion used. At zero-degree inclination, we observe that Earth-crossing orbits generally have shorter decoherence lifetimes, on the order of tens to hundreds of thousands of years. As the inclination increases to 10 and 20 degrees, these lifetimes extend but still remain within this same general range. These results are consistent with those found by \citet{pauls2005decoherence}, indicating shorter lifetimes for streams on Earth-crossing orbits. Furthermore, the influence of Jupiter is apparent, with streams at higher inclinations or those closer to Jupiter's orbit experiencing more rapid decoherence due to the planet's gravitational perturbations. This indicates that while tidal disruptions may create clusters, the chaotic nature of these orbits results in a rapid loss of coherence, emphasizing the ephemeral nature of such streams in near-Earth space.

% talk about as function of D-value and why
There also appears to be some noticeable variation between the orbital similarity discriminants used in this study. While the overall trends in decoherence lifetimes as a function of semi-major axis, eccentricity, and inclination remain consistent across the different discriminants, the absolute values of the lifetimes do show some variation. For instance, the $D_H$ and $D_{SH}$ discriminants exhibit nearly identical distributions, with $D_{SH}$ showing slightly, though very minutely, longer decoherence lifetimes. In contrast, the $D'$ discriminant consistently shows longer decoherence lifetimes, a difference that is likely attributable to its stronger dependence on eccentricity \citep{jopek1993remarks}. This effect is evident in Fig.~ \ref{fig:0deg_decoherence_maps}, \ref{fig:10deg_decoherence_maps}, and \ref{fig:20deg_decoherence_maps}, where the decoherence lifetimes associated with $D'$ are slightly shifted towards larger values compared to those of $D_H$ and $D_{SH}$ regardless of the inclination. $D_H$ and $D_{SH}$, which are more dependent on perihelion distance than $D'$, show more rapid changes in orbital similarity due to these encounters, leading to shorter decoherence lifetimes. However, despite these differences, the overall trends remain consistent, and the decoherence lifetimes for Earth-crossing meteoroid streams across all three discriminants are generally within the 10-100\,kyr range with evolved orbits (semi-major axis no longer in range of the main-belt) capable of longer decoherence lifetimes.

\section{Discussion}\label{sec:discuss}

Using a kernel density estimation (KDE) method modified from \citet{shober2024generalizable}, we evaluated the statistical significance of orbital pairings within several meteoroid and near-Earth object (NEO) datasets. This method allowed us to assess over \(3\times10^{11}\) possible pairs amongst terrestrial impact datasets and NEOs, supplemented by Monte Carlo simulations to rigorously quantify the significance of these pairings, as outlined in section \ref{sec:method}.

Our analysis revealed that within the meteorite fall database, comprising 50 meteorites with well-documented orbits, there is no conclusive evidence of statistically significant pairings. Furthermore, when compared to the NEO clusters, only 7 fireball events of the 616 possible 1\,g falls observed by the GFO, FRIPON, and the EFN were identified to satisfy the DBSCAN clustering $\sigma$ and N$_{min}$ constraints as used for Fig.~\ref{fig:NEO_clusters}. None of the 50 meteorite falls or the 310 CNEOS impacts were identified to be clustered. The cumulative distribution of D-values (D$_{SH}$, D', and D$_H$) across these datasets was linear, which aligns with random chance associations rather than genuine clustering. Also, more importantly, the cumulative D-value distributions corresponded well with the number expected due to chance association. Even when considering clustering with the NEO population, only $\sim$1.2\% are identified to be similar to identified NEO clusters with 5 or more members. Despite numerous studies attempting to link specific meteorite falls with NEOs via orbital similarity, our findings suggest that all these pairings lack statistical significance. The linearity observed in the cumulative distribution of D-values strongly indicates that any observed orbital similarities between meteorites and NEOs are coincidental and not indicative of actual streams. 

In contrast, our analysis confirms the presence of statistically significant clusters within the NEO population, corroborating the findings of \citet{jopek2020orbital}. However, our results suggest a more conservative estimate of the number of such clusters, identifying 12 statistically significant clusters compared to the 15 reported by \citet{jopek2020orbital}. Despite this difference in numbers, the locations of these clusters -- primarily around perihelion distances near Earth's orbit and in low-inclination orbits -- are consistent with previous studies by \citet{schunova2014properties} and the recent findings by \citet{granvik2024tidal} for evidence of tidal disruption in near-Earth space. The clustering of NEOs in these regions aligns with the hypothesis that tidal disruptions of NEOs, particularly during close encounters with Earth and Venus, produce a statistically significant excess of objects in these orbits. 

It is essential to note that this does not necessarily mean that no meteorites are being produced from this tidal disruption mechanism. Quite the contrary, in a very recent work, \citet{shober_carbonaceous} argues that the release of CI/CM chondrites from an immediate precursor body already in near-Earth space would much better explain the CRE ages, petrographic features, mixture of irradiated grains, and isotopic geochemistry in the samples. There may be many reasons why no meteorite with orbits and fireball observations are above the detectable limit. Firstly, the tidal disruption mechanism is not so dominant in near-Earth space that it produces a significant proportion of the debris. According to \citet{jopek2020orbital}, only 4.7\% of the NEOs are identified in clusters, whereas we find this value to be likely lower, possibly below 1\%. In any case, these minor streams require significant amounts of meteorite-dropping fireball observations to be detectable. Thus, one likely explanation is that we simply that we don't have enough observations yet. 

Another reason is that even on an orbit with a perihelion distance near the Earth, the impact timescales of the meteoroids produced from these tidal disruptions or other mechanisms (e.g., meteoroid impacts; \citealp{turner2021carbonaceous,shober_carbonaceous}) are just too prolonged. The inverse of typical impact frequencies (10$^{8}$-10$^{10}$\,yrs; \citealp{bottke1994collisional}) with the Earth are significantly longer than the decoherence lifetimes of any possible stream. Also, the tidal disruption frequency, estimated to be once every $\sim$\,2,500\,years \citep{schunova2014properties}, are 10-1000x shorter than the dynamical lifetimes in the inner solar system \citep{gladman1997dynamical}. Thus, if tidal disruption is a consistent mechanism of meteoroid and small asteroid production near the Earth, due to the inherent shortness of stream decoherence, a large majority of the debris being detected telescopically or by fireball networks would necessarily not be expected to be found in distinct streams. This could likely be why \citet{jopek2020orbital} found many NEOs, generally clusters of q\,$\sim$\,1\,au orbits, an observation we confirm here. Many of these NEOs do not fulfill our statistical significance test; however, the general clusters on these orbits could be representative of older, no longer coherent streams. In addition, the decoherence lifetimes of streams (typically 10$^{4}$-10$^{5}$ years), tend to be significantly shorter than the cosmic-ray exposure (CRE) ages of meteorites  (Fig.~\ref{fig:0deg_decoherence_maps},\ref{fig:10deg_decoherence_maps},\ref{fig:20deg_decoherence_maps}). Thus, any proposed meteorite linkages in the future must also have a shorter CRE age than the expected decoherence lifetime of a stream on a nearby orbit. This is precisely the reasoning of \citet{shober_carbonaceous}, which found that CI/CM meteorites Sutter`s Mill, Flensburg, and Winchcombe have CRE ages as younger or significantly younger than the average CRE age for debris released already on an Earth-crossing orbit. 

The only way to have longer CRE ages and still be ejected from an NEO would be to collect much of the CRE age on the surface of the NEO. This would leave a ``complex'' exposure signature in the irradiation history \citep{wieler2001cosmic}. One of the primary reasons the Příbram/Neuschwanstein pair was disregarded in the first place was that the CRE ages and meteorite types were significantly different, Příbram being an H5 with a CRE age of $\sim$12\,Myrs and Neuschwanstein being an EL6 with a CRE age between 43-51\,Myrs  \citep{stauffer1962multiple,zipfel2010mineralogy,pauls2005decoherence}. However, as observed with the fall of Almahatta Sitta (asteroid 2008 TC3), this original logic that meteorites originating from the same meteoroid impact necessarily need to share a common composition and irradiation history is understood to no longer be a requirement \citep{bischoff2010asteroid}. Thus, a pair like Příbram/Neuschwanstein could be related and generated during a disruption of a polymict near-Earth rubble pile \citep{toth2011tidal,granvik2024tidal}, assuming a complex exposure is consistent with the cosmogenic radionuclide and noble gas measurements \citep{meier2022pribram}. Nevertheless, since the sole reason for the association is orbital similarity, without some additional constraint based on the meteorites themselves, this pair is unfortunately incapable of confirmation. For bodies separated in space for more than a few centuries, orbital similarity discriminants can only detect streams of many objects, not pairs. Even when a stream is well identified, there will always be some sporadic component erroneously included \citep{shober2024generalizable}, but without some other constraint, you cannot be sure which objects expressly are the false positives, only the proportion of the sample. 

\section{Conclusion}
In this study, we revisited the meteoroid and NEO stream decoherence concept and critically assessed the limits of orbital similarity measures in stream identification. Our findings emphasize the transient nature of near-Earth streams and challenge the traditional reliance on orbital similarity discriminants for individual pair associations. The primary results of this research are summarized as follows:

\begin{itemize}
    \item Using a Kernel Density Estimation (KDE) method \citep{shober2024generalizable}, we evaluated the statistical significance of orbital pairings within various datasets, including 50 meteorite falls with orbits, 616 potential $>$1\,g meteorite-dropping fireballs (of which 350 are likely $>$50\,g), 310 impact detected by US government sensors, and 35,012 NEOs. Our analysis revealed no statistically significant streams in the meteorite fall, USG impact, or fireball datasets. The cumulative D-value distributions for these datasets were consistent with random chance, with no observed clustering beyond what would be expected randomly. 

    \item Conversely, we identified 11 statistically significant clusters within the NEA population and one of 70 fragments associated with the disintegration of comet 73P/Schwassmann-Wachmann. All clusters are found in orbits with perihelion distances near 1\,au. These clusters, identified using the DBSCAN algorithm with $D_H$ values below 0.03, likely result from tidal disruption events, consistent with the findings of \citet{granvik2024tidal}. These clusters suggest that while streams may rapidly decohere, the process generating them is frequent enough to maintain small identifiable NEO families.

    \item Within this study, we utilized traditional orbit similarity discriminants; however, these orbital discriminates are simplistic and are not ideal for clustering orbital data in all cases. The orbital evolution of streams in the solar system depends on the locations of the streams and the primary forces driving this evolution (which could be size-dependent). The weightings of the orbital parameters within orbital similarity discriminants do not reflect this. Thus, new methods that take into account this orbital evolution variation should be developed to more accurately identify and characterize groupings of comets, asteroids, and meteoroids. There is a statistically significant amount of similarity within the NEO population, but the NEO families identified here using current orbital similarity discriminants should be re-evaluated using a method that takes into account the local dynamics of streams. 

    \item Furthermore, none of the fireball data examined (616 possible droppers + 50 meteorite falls) fulfilled the DBSCAN requirements to be members of any of the 12 NEO clusters identified. Even when the minimum cluster size was lowered from five to three members, no impact data was still classified as being member of one of the identified NEO clusters. 

    \item Lyapunov lifetime analysis demonstrated that orbits in near-Earth space, particularly those with semi-major axes between 0.5 and 3.5 AU and eccentricities between 0.2 and 0.9, exhibit chaotic behavior with characteristic lifetimes as short as 60 to 200 years (Fig.~\ref{fig:lyapunov_maps}). These short timescales underscore the rapid divergence of initially similar orbits, rendering long-term orbital similarity measures unreliable for individual pair identification.

    \item The decoherence lifetimes of meteoroid streams, as defined as the time required for a stream to lose 95\% of its original members, ranged from $10^4$ to $10^5$ years for Earth-crossing orbits. This aligns with previous studies such as \citet{pauls2005decoherence}, indicating that streams on these orbits lose coherence on timescales significantly shorter than most meteorites' cosmic ray exposure ages. Thus, most meteorites would require the presence of some ``complex'' cosmic-ray exposure signature to be considered as possibly ejected recently by an immediate-precursor body in near-Earth space. 
\end{itemize}

Our results highlight a significant limitation in using orbital similarity measures: they are not reliable for identifying specific parent bodies of meteorites. The Lyapunov characteristic timescales in near-Earth space are only a few hundred years, after which orbits diverge chaotically. This rapid separation in nearby orbits means that any association between a specific fireball, meteorite, and asteroid, based solely on orbital similarity, is likely coincidental rather than indicative of a true physical connection. Orbital similarity can provide a potential avenue to identify objects with a possible `genetic' link; however, more constraints are always needed to confirm this. While these metrics help identify streams, they cannot confirm specific body associations alone. Despite the evidence that meteorites can be and are ejected from immediate precursor bodies already within near-Earth space \citep{granvik2024tidal,shober_carbonaceous}, these streams become decoherent rapidly. Identifying such streams using similarity alone within the meteorite and fireball populations will require a significantly larger dataset than we currently have. We would likely need tens of thousands of meteorite-dropping fireball observations to detect these small ephemeral streams with enough statistical significance to separate them from random associations with the sporadic complex if the stream-sporadic ratio is similar to NEOs. Only with such extensive data could we begin to discern small streams resulting from tidal disruptions or other mechanisms, such as meteoroid impacts. In any case, the notion of confidently linking a single fireball to a single NEO through orbital similarity remains fundamentally flawed and unattainable due to the inherent limitations imposed by the chaotic nature of orbits in near-Earth space.

% future work ! examining decoherence timescales for cometary meteor showers (include non-gravs), another way to find streams... similarity plus other things, but not just similarity. Can confirm streams... not individual links!

\section*{Acknowledgement}\label{sec:ack}

This project has received funding from the European Union’s Horizon 2020 research and innovation programme under the Marie Skłodowska-Curie grant agreement No945298 ParisRegionFP. 

The Global Fireball Observatory and data pipeline is enabled by the support of the Australian Research Council (DP230100301, LE170100106).

FRIPON was initiated by funding from ANR (grant N.13-BS05-0009-03), carried by the Paris Observatory, Muséum National d’Histoire Naturelle, Paris-Saclay University and Institut Pythéas (LAM-CEREGE). VigieCiel was part of the 65 Millions d’Observateurs project, carried by the Muséum National d’Histoire Naturelle and funded by the French Investissements d’Avenir program. FRIPON data are hosted and processed at Institut Pythéas SIP (Service Informatique Pythéas), and a mirror is hosted at IMCCE (Institut de MécaniqueCéleste et de Calcul des Éphémérides / Paris Observatory). 

This work was granted access to the HPC resources of MesoPSL financed by the Region Ile de France and the project Equip@Meso (reference ANR-10-EQPX-29-01) of the programme Investissements d’Avenir superv ised by the Agence Nationale pour la Recherche. 

This research used Astropy, a community-developed core Python package for Astronomy \citep{robitaille2013astropy}. Simulations in this paper used the REBOUND software package\footnote{\url{http://github.com/hannorein/REBOUND}} \citep{rebound2012A&A...537A.128R}. 

%The authors declare no competing interests. All data needed to evaluate the conclusions in the paper are present in the paper and/or at \url{https://doi.org/10.5281/zenodo.4710556}.

\bibliographystyle{aa} 
\bibliography{main.bib}

\end{document}